\documentclass[journal=jctcce]{achemso}

\usepackage[T1]{fontenc} 
\usepackage{amsmath}
\usepackage{amsfonts}
\usepackage{amsthm}
\usepackage{mathtools}

\usepackage[%
  colorlinks=true,
  allcolors=blue
]{hyperref}

\usepackage{natbib}
\usepackage{textcomp}

\DeclarePairedDelimiter\ppar{(}{)}              
\DeclarePairedDelimiter\pabs{\lvert}{\rvert}    
\DeclarePairedDelimiter\pnrm{\lVert}{\rVert}    
\DeclarePairedDelimiter\pbkt{[}{]}              
\DeclarePairedDelimiter\pset{\{}{\}}            

\newcommand{\rfig}[1]{Figure~\ref{#1}}

\newcommand{\rsct}[1]{Section~\ref{#1}}

\newcommand{\req}[1]{eq~\ref{#1}}

\newcommand{\cond}{\left.\right\vert}
\newcommand{\us}{{\bf s}}
\newcommand{\uv}{{\bf v}}
\newcommand{\uz}{{\bf z}}

\newcommand{\ux}{{\bf x}}

\newcommand{\uq}{{\bf q}}
\newcommand{\uM}{{\bf M}}
\newcommand{\uMH}{\mathbf{M}_{\mathrm{H}}}
\newcommand{\uT}{{\bf T}}
\newcommand{\uP}{{\bf P}}
\newcommand{\dub}{{\rm ub}}
\newcommand{\db}{{\rm b}}
\newcommand{\tus}{\tilde{\bf s}}
\newcommand{\sus}{\mathbf{s}^\ast}
\newcommand{\xus}{\mathbf{x}^\ast}

\newcommand{\detm}[1]{\operatorname{\pabs{#1}}}

\author{Yanbin Wang}
\affiliation[Purdue University]{Department of Chemistry, Purdue University, West Lafayette}

\author{Jakub Rydzewski}
\affiliation[NCU]{Institute of Physics, Faculty of Physics, Astronomy and Informatics, Nicolaus Copernicus University, Grudziadzka 5, 87-100 Toru\'{n}, Poland}

\author{Ming Chen}
\affiliation[Purdue University]{Department of Chemistry, Purdue University, West Lafayette}
\email{chen4116@purdue.edu}

\title{Constructing Generalized Sample Transition Probabilities with Biased Simulations}

\begin{document}

\begin{tocentry}
    \centering
    \includegraphics{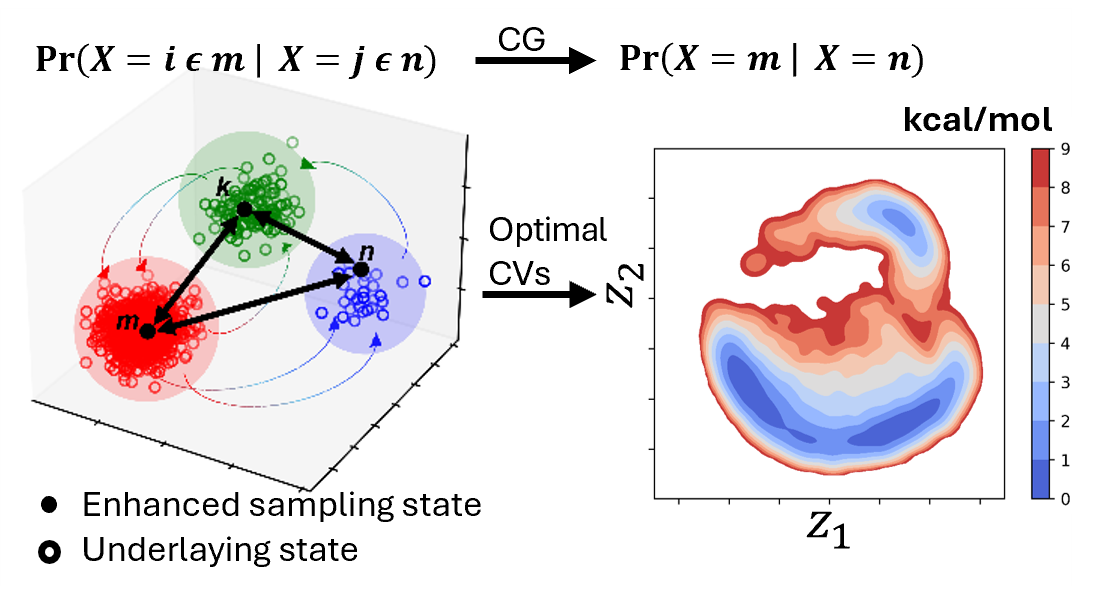}
\end{tocentry}

\begin{abstract}

In molecular dynamics (MD) simulations, accessing transition probabilities between states is crucial for understanding kinetic information, such as reaction paths and rates. However, standard MD simulations are hindered by the capacity to visit the states of interest, prompting the use of enhanced sampling to accelerate the process. Unfortunately, biased simulations alter the inherent probability distributions, making kinetic computations using techniques such as diffusion maps challenging. Here, we use a coarse-grained Markov chain to estimate the intrinsic pairwise transition probabilities between states sampled from a biased distribution. Our method, which we call the generalized sample transition probability (GSTP), can recover transition probabilities without relying on an underlying stochastic process and specifying the form of the kernel function, which is necessary for the diffusion map method. The proposed algorithm is validated on model systems such as a harmonic oscillator, alanine dipeptide in vacuum, and met-enkephalin in solvent. The results demonstrate that GSTP effectively recovers the unbiased eigenvalues and eigenstates from biased data. GSTP provides a general framework for analyzing kinetic information in complex systems, where biased simulations are necessary to access longer timescales.
\end{abstract}

\section{Introduction}

Molecular dynamics (MD) simulations offer insight into physical and chemical processes at atomic resolution. However, the timescales of rare events, such as protein folding~\cite{lindorff2011fast}, crystallization~\cite{neha2022collective}, nucleation~\cite{beyerle2023recent}, catalysis~\cite{piccini2022ab}, or molecular recognition~\cite{baron2013molecular,rydzewski2017ligand}, are much larger than those of atomic vibrations. The MD sampling problem requires enhanced sampling techniques to explore the diverse states of a system efficiently~\cite{valsson2016enhancing,bussi2020using}. Enhanced sampling methods, including metadynamics~\cite{laio2002escaping,barducci2008well}, umbrella sampling~\cite{torrie1977nonphysical,Kastner2011umbrella}, temperature accelerated molecular dynamics (TAMD)~\cite{voter2000temperature}, and adaptive biasing force~\cite{darve2008adaptive}, are often used to solve this problem and drive MD simulations to explore complex free energy landscapes (FELs).

Many enhanced sampling methods rely on biasing the probability of a few variables, called collective variables (CVs), reaction coordinates, or order parameters. CVs are variables that can describe slow modes in structural dynamics and are expected to map conformational changes on a low-dimensional FEL~\cite{fiorin2013using,noe2017collective,yu2014order,tse2019affordable,bidon2013membrane,chipot2005exploring,bonhenry2018effects,samanta2014microscopic,yu2011temperature,abrams2010large,barducci2011metadynamics,laio2002escaping,laio2008metadynamics,valsson2016enhancing,singh2012density,pietrucci2009collective,mendels2018collective,rogal2021reaction,gokdemir2024machine}. Therefore, algorithms for enhancing the barrier crossing in an FEL can enhance important transitions, leading to efficient exploration of various conformations. Conventionally, CV design is often based on our intuition and understanding of simulated processes~\cite{bussi2015free,valsson2016enhancing,rogal2021reaction,gokdemir2024machine}. Although such CVs have a physical interpretation, it is highly challenging to map all important conformations onto the FEL. Suboptimal CVs can leave energy barriers in the orthogonal space, i.e., multiple conformations are mapped onto the same minimum. Therefore, suboptimal CVs can impede sampling efficiency or even its accuracy~\cite{zheng2008random,zhang2018unfolding}. Recently, machine learning has been used to learn CVs from MD simulation data~\cite{bonati2021deep,trizio2021enhanced,sidky2020machine,chen2021collective,bonati2023unified,zhang2018unfolding,zhang2018unfolding,bonati2020data,sipka2023constructing,rydzewski2023spectral,rydzewski2024tse,bonati2020data,dorfer2015deep,lemke2019encodermap,schoberl2019predictive,hernandez2018variational,tajs2025neuraltsne}. 

There are various approaches to estimating the kinetics from MD simulations. In unbiased MD simulations, kinetics can be modeled by transition path sampling~\cite{bolhuis2002transition}, Markov state models~\cite{chodera2014markov}, or milestoning~\cite{elber2020milestoning}. Kinetic information from unbiased MD simulations has been applied to learn CVs \cite{tiwary2016spectral,tsai2020learning,rydzewski2023spectral,rydzewski2024learning,rydzewski2024tse}. Since CV training is often required directly from biased enhanced sampling data~\cite{mccarty2017variational,belkacemi2021chasing,rydzewski2021multiscale,rydzewski2022reweighted}, unbiasing kinetics is required. However, unbiasing is challenging because biased sampling, which aims to accurately reconstruct thermodynamics, fails to retain the correct dynamics. Some methods are specifically designed for metadynamics that can unbias path probabilities. For example, a change of variables in time can asymptotically correct the accelerated time scale in well-tempered metadynamics~\cite{mccarty2017variational}. Recently, the Girsanov formula has been applied to calculate the exact path reweighting factors for metadynamics~\cite{donati2017girsanov}. However, a generalized technique for unbiasing transition probabilities is not available for any enhanced sampling method. 

Due to the challenges of unbiasing transition probabilities, the so-called spatial techniques provide an alternative approach to approximate kinetics with thermodynamical information~\cite{gokdemir2024machine}. Such techniques also provide a way to construct CVs from biased enhanced sampling data. One such example is diffusion map~\cite{coifman2006diffusion,coifman2008diffusion} (DM), which uses equilibrium probability to build a Markov chain on configurations sampled from a diffusion process. This Markov chain can be used to approximate the generator for the diffusion process~\cite{coifman2006diffusion}. DM itself can be used to define CVs by approximating them using eigenvectors of a transition matrix~\cite{coifman2008diffusion,rohrdanz2011determination}. DM has also been employed in other CV-training frameworks~\cite{zhang2018unfolding,rydzewski2022reweighted,rydzewski2023selecting}. A generalization of a Gaussian kernel used in DM leads to the definition of Mahalanobis DM~\cite{singer2008non,singer2009detecting} (MDM). Recently, MDM has been developed to address the diffusion process of states in the feature space~\cite{evans2022computing,evans2023computing}.

Both DM and MDM can be constructed from biased sampling simulations~\cite{banisch2020diffusion,trstanova2020local,evans2022computing,rydzewski2022reweighted,rydzewski2023selecting}. As these methods are based on diffusion processes, unbiasing DM and MDM is based on preserving the correct semigroup with respect to a diffusion process of unbiased simulation~\cite{evans2023computing}. The idea of DM and MDM can be generalized to define generalized pairwise transition probabilities. Although they share a similar form to that of DM or MDM, they are not derived from any stochastic differential equation (SDE). Therefore, the unbiasing formalism published in \citep{evans2022computing,evans2023computing} is designed for specific cases.

In this work, we will present a generalized formalism of unbiasing pairwise transition probabilities without relying on a diffusion process. We call it the generalized sample transition probability (GSTP). The idea is to use a set of biased samples to partition the Cartesian or feature space into ``cells'' so that the pairwise probability of a pair of biased samples can be viewed as a coarse-grained probability defined between two corresponding cells. The proposed method generates results for DM and MDM that are consistent with previous studies. We will also demonstrate that the method can correctly unbias pairwise transition probabilities with different selections of kernel functions that are not suitable for DM or MDM. 

The paper is organized as follows. We will first briefly introduce the background of DM and MDM. After introducing the existing theory, we will present our proof of the proposed approach, followed by numerical justifications. We will show the testing results on a diffusion process with harmonic potential with which analytical results of the generator's eigenvalues and eigenvectors are known. We will further test the unbiasing results with alanine dipeptide and met-enkephalin in explicit solvent and show that the free energy profiles of the unbiased enhanced sampling samples align with those of unbiased samples, irrespective of the choice of kernel type or enhanced sampling method.

We conclude that GSTP is a general and robust method for revealing the pairwise transition probability distribution from biased simulations. This work provides a simple but effective approach to capture the kinetic information of a complex system that often relies on enhanced sampling to explore the states of interest.

\section{Background}
\subsection{Dynamics in the Feature Space}
%

Consider a system of $N$ atoms with Cartesian coordinates $\ux=(x_1,x_2,\cdots,x_{3N})$ whose dynamics at temperature $T$ evolves in a potential energy function $U(\ux)$. In the $NVT$ ensemble, the equilibrium probability distribution of the system is the Boltzmann distribution: 
\begin{equation}
    \rho(\ux)=\frac{1}{Z}e^{-\beta U(\ux)}
\end{equation}
where $\beta=1/k_{\mathrm{B}}T$ is the inverse temperature and $Z$ is the configuration integral. To simplify the representation of the system, we introduce $n$ functions of Cartesian coordinates $\uq=(q_1(\ux),\cdots,q_n(\ux))$. As they may not correspond to the slow modes of the system, we refer to them as features. The marginal probability at $\uq(\ux)=\us$ is the following:
\begin{equation}
    \rho(\us)=\frac{1}{Z}\int\mathrm{d}\ux e^{-\beta U(\ux)}\delta(\uq(\ux)-\us)
    \label{eq:prob-s}
\end{equation}
or equivalently:
\begin{equation}
    \rho(\us)=\frac{1}{Z_s} e^{-\beta A(\us)},
    \label{eq:prob-s-equiv}
\end{equation}
where $A(\us)$ is the FEL and $Z_s$ is the partition function in the feature space. To clarify, we define features as variables that do not necessarily describe slow modes, in contrast to CVs.

Many methods for generating states that conform to the correct probability distribution at equilibrium are available, with MD simulations being a widely used approach. By treating MD as a state‑generator for physical or chemical systems, we can sample configurations that, over sufficiently long simulations, yield reliable thermodynamic properties.

Suppose that a random configuration is chosen as our initial state. This initial state, whether defined by Cartesian coordinates or other features, in the next time step will follow a distribution that can be described using an SDE, such as the overdamped Langevin equation in the $NVT$ ensemble. To model the time evolution of the system, we will use the following diffusion equations. In the coordinate space, we have:
\begin{equation}
    \mathrm{d}\ux = -\nabla U(\ux) \mathrm{d}t+\sqrt{2 \beta^{-1}}\mathrm{d}\mathbf{w},
\label{eq:ger_co}
\end{equation}
where $\mathrm{d}\mathbf{w}$ is the Brownian motion. The infinitesimal generator of this diffusion process is:
\begin{equation}
    \mathcal{L} = -\nabla U(\ux)\cdot\nabla+\beta^{-1}\nabla^2\;\;\ldotp
\label{eq:ger_co_2}
\end{equation}
As features are defined as functions of Cartesian coordinates, the states generated in the feature space can be described by applying Ito's lemma. Thus, \req{eq:ger_co} results in a diffusion equation in the feature space~\cite{evans2023computing,evans2022computing}:
\begin{equation}
\mathrm{d}\us = \ppar[\big]{-\uM(\us)\nabla A(\us)+\beta^{-1}\nabla\cdot\uM(\us)}\mathrm{d}t
+\sqrt{2\beta^{-1}\uM(\us)}\mathrm{d}\mathbf{w}
\label{eq:qsde}
\end{equation}
where $\uM(\us)$ is the diffusion matrix:
\begin{equation}
\uM(\us) = e^{\beta A(\us)}\int\mathrm{d}\ux\,
J(\ux)J^\top(\ux)\frac{1}{Z}e^{-\beta U(\ux)}\delta(\uq(\ux)-\us),
\label{eq:diffmat}
\end{equation}
in which $J(\ux)$ is the Jacobian matrix:
\begin{equation}
J_{\alpha l}(\ux)=\frac{\partial q_\alpha}{\partial x_l},
\end{equation}
and $\nabla\cdot\uM(\us)$ is defined as a vector whose element $\alpha$ is $\sum_\beta\frac{\partial M_{\alpha\beta}(\us)}{\partial s_\beta}$. The generator of the diffusion process described by \req{eq:qsde} is:
\begin{equation}
    \mathcal{L} = \ppar[\big]{-\uM(\us)\nabla A(\us)+\beta^{-1}\nabla\cdot\uM(\us)}\cdot\nabla + \beta^{-1}\nabla\cdot\ppar[\big]{\uM(\us)\nabla}.
\end{equation}
More details about this equation can be found in the work by Maragliano and Vanden-Eijnden~\cite{maragliano2006string}.

\subsection{Diffusion Map}
\label{sec:dms}
The eigenvectors of the generator can inherently represent the slow modes of the studied system. For example, methods such as Markov state models have been used to model the eigenvectors of generators~\cite{chodera2014markov}. However, their construction requires limiting the perturbation in a simulation and employing enhanced sampling methods in modeling the eigenvectors of the generator is challenging~\cite{keller2024dynamical}. 

In contrast, DM and MDM approximate the eigenvectors of $\mathcal{L}$ with the equilibrium probability distribution. The idea is to build a Markov chain $T_{ij}$ on sampled configurations. This Markov chain is then able to approximate the eigenvector of $\mathcal{L}$ at the sample points. We will use the MDM algorithm to demonstrate how to obtain kinetic information with the equilibrium probability distribution. One of the advantages of MDM is that it is built solely from datasets in thermodynamic equilibrium. The algorithm for constructing a Markov chain with MDM from a dataset of $N$ samples $\{\us_i\}$ is the following~\cite{evans2023computing}:

\begin{enumerate}
    \item We construct a kernel by evaluating similarities between samples $\us_i$ and $\us_j$ using a Gaussian form, i.e., $K_{ij}=G_s(\us_i,\us_j)$~\cite{evans2023computing,evans2022computing}: 
    \begin{equation}
        G_s(\us_i,\us_j) = \exp\ppar*{-\frac{1}{4\sigma^2}(\us_i-\us_j)^\top(\uM^{-1}(\us_i)+
            \uM^{-1}(\us_j))(\us_i-\us_j)},
    \label{eq:feature_kernel}
    \end{equation}
    where $\uM(\us_i)$ is the diffusion tensor and $\sigma$ is a bandwidth.

    \item We estimate a prototype transition matrix by:
    \begin{equation}
        D_{ij} = \frac{G_s(\us_i,\us_j)}{\sqrt{\rho_M(\us_i)}\sqrt{\rho_M(\us_j)}},
    \end{equation}
    where $\rho_M(\us_i)\propto \int\mathrm{d}\us'G_s(\us,\us')\rho(\us')$.

    \item The matrix $D_{ij}$ is normalized to construct a transition matrix $\uT^\dub$:
    \begin{equation}
        T^{\dub}_{ij}=\frac{D_{ij}}{\sum_k D_{ik}},
        \label{eq:Tij}
    \end{equation}
    where the transition probability is $T^\dub_{ij}=P(\us(t+1)=\us_j \cond \us(t)=\us_i)$. It has been proved that, in the limit of $N\to\infty$ and $\sigma\to 0$,  the estimate $L_{ij} = \frac{T^\dub_{ij}-\delta_{ij}}{\sigma^2} $ weakly converges to $\frac{\beta}{2}\mathcal{L}$~\cite{evans2022computing,evans2023computing}, where $\delta$ is the Kronecker delta function.

    \item Finally, the diffusion coordinates (eigenvectors of the transition matrix) can be defined as CVs~\cite{coifman2006diffusion,coifman2008diffusion}. They are calculated by solving an $N \times N$ eigendecomposition problem $\uT^\dub z_i = \lambda_i z_i$:
    \begin{equation}
        \uz = \ppar*{z_0,z_1, z_2, \cdots, z_p},
    \end{equation}
    where the eigenvalues are sorted in non-decreasing order $\lambda_0 = 1 \ge \lambda_i \ge \cdots$ and $z_i$ are the corresponding eigenvectors. The eigenvalues $\lambda_i$ are related to the eigenvalues of $\mathcal{L}$, $\epsilon_i\approx \tilde{\epsilon}_i\equiv \frac{2(\lambda_i-1)}{\beta\sigma^2}$. In the following context, we refer to $\tilde{\epsilon}$ as scaled eigenvalues. The eigenvector $z_0$ is the equilibrium density. The dimension of $\uz$ is denoted as $p$ and marks the position of the spectral gap in the eigenspectrum, i.e., $\lambda_p \gg \lambda_{p+1}$.
    
\end{enumerate}

In the special case where the kernel is built on a dataset in the coordinate space, i.e., $\uq \equiv \ux$, \req{eq:feature_kernel} reduces to an isotropic Gaussian kernel:
\begin{equation}
    G(\ux_i,\ux_j)=\frac{1}{(2\pi\sigma^2)^{\frac{d}{2}}}
        \exp\left(-\frac{\|\ux_i-\ux_j\|^2}{2\sigma^2}\right),
    \label{eq:kernel}
\end{equation}
where $d$ is the dimension of $\ux$. Then, MDM becomes DM and follows the same process of building the transition matrix $\uT^\dub$.

\subsection{Reweighting Transition Probabilities}
Although MDM construction with the above protocol does not require explicitly calculating kinetic quantities such as correlation functions, the algorithm needs samples generated by unbiased simulations. This requirement limits the application of DM and MDM to datasets obtained only from unbiased simulations. The probability distributions $p(\ux)$ and $p(\us)$ can be generated by reweighting enhanced sampling simulations~\cite{tiwary2015time,liu2024unbiasing}. However, further changes are needed in the transition matrix to ensure that it is consistent with the one calculated from unbiased data~\cite{evans2022computing,evans2023computing}. 

In the following, we will use notation such that the dataset of unbiased samples in the coordinate and feature spaces is given as $\pset*{\ux_i}~\mathrm{and}~\pset*{\us_i}~\mathrm{for}~i=1,\dots,N$, respectively. To denote that samples are generated from a biased distribution, we will mark them with an asterisk:
\begin{equation}
    \pset[\big]{(\xus_i, \omega_i)}~\mathrm{and}~\pset[\big]{(\sus_i, \omega_i)}~\mathrm{for}~i=1,\dots,N,
\end{equation}
where $\omega$ is the unbiasing weight.
For a set of biased samples in the feature space with weights, the transition matrix in MDM is:
\begin{equation}
    T^\db_{mn} = \frac{\frac{\omega_n}{\sqrt{\rho(\sus_n)}} G_s(\sus_m,\sus_n) \detm{\uM(\sus_n)} ^{-1/4}}
{\sum_l \frac{\omega_l}{\sqrt{\rho(\sus_l)}} G_s(\sus_m,\sus_l)\detm{\uM(\sus_l)}^{-1/4}},
\label{eq:hete_gauss}
\end{equation}
where $\detm{\uM}$ denotes the determinant of the diffusion matrix $\uM$. As a special case, the transition matrix for DM becomes:
\begin{equation}
    T^\db_{mn}= \frac{\frac{\omega_n}{\sqrt{\rho(\mathbf{x}^\ast_n)}} G(\xus_m,\xus_n)}{\sum_l \frac{\omega_l}{\sqrt{\rho(\xus_l)}} G(\xus_m,\xus_l)
}\ldotp
\label{eq:unbias-mdm}
\end{equation}
The above formulas are limited to DM and MDM with kernels defined in \req{eq:kernel} and \req{eq:feature_kernel}, respectively. We want to emphasize that \req{eq:hete_gauss} is equivalent to the unbiasing formula in \citet{evans2023computing} (see Supporting Information for the proof of \req{eq:hete_gauss}). Unlike the MDM formula presented in \rsct{sec:dms}, \req{eq:unbias-mdm} uses $\rho$ instead of $\rho_M$, which allows the potential usage of other methods, such as diffusion models or normalizing flows~\cite{liu2024unbiasing} to calculate $\rho$.

\section{Generalized Sample Transition Probabilities}
\label{sec:GSTP}
In applications, the kernel function is sometimes changed to another form instead of Gaussians~\cite{rydzewski2023manifold}, or the distance between samples is given by another metric to measure the similarity of configurations. Such modifications will be described as generalizations of the transition probabilities in DM and MDM, and we will refer to them as generalized sample transition probabilities (GSTPs), denoted as $\uP$. Importantly, we will show that even if GSTPs are not defined on the basis of the diffusion process, the unbiasing formulas are surprisingly similar to those of DM and MDM. In this Section, we will discuss a new way to obtain an unbiasing formula for GSTPs. We will first define GSTP with unbiased data. 
We will propose a method that uses coarse-graining to construct GSTPs from samples generated from a biased distribution. We will show that the unbiasing protocol leads to eigenvalues and eigenvectors of $\uP$ that are consistent with those of the unbiased samples.

\subsection{Case I: Unbiased Data}
\label{sec:gstp-unbiased}
Before discussing how to calculate the GSTPs, 
we will estimate $\pi(\us_i)$, the long‑time probability of finding the system in state $\us_i$ when the Markov chain has reached equilibrium, without the need to employ the Gaussian kernel. Note that $\pi(\us_i)$ needs to be evaluated with unbiased samples and the transition matrix $T_{ij}^\dub$ from \req{eq:Tij} will be used. In DM or MDM, the right eigenvectors are components of the spectral decomposition of the transition matrix that represents the diffusion process. For a detailed derivation, we refer to Supporting Information.

Similarly as before (see \rsct{sec:dms}), a prototype generalized transition matrix is constructed as:
\begin{equation}
    D_{ij} = \frac{K_s(\us_i,\us_j)}{\sqrt{\rho_M(\us_i)}\sqrt{\rho_M(\us_j)}},
\end{equation}
where $\rho_M(\us_i)$ can be calculated as the expectation of the probability of sample $\us_i$ in the feature space $\rho(\us_i)$ (using a standard Gaussian kernel) weighted by a generalized kernel function $K_s(\us_i,\us_j)$. Although $K_s(\us_i,\us_j)$ does not have to be a Gaussian kernel, it needs to satisfy certain conditions. Specifically, we assume $K_s(\us_i,\us_j)=k(d^2(\us_i,\us_j);\sigma)$ is a non-negative function with a single parameter $\sigma$ and that $d^2(\us_i,\us_j)$ measures the similarity between $\us_i$ and $\us_j$. That is:
\begin{equation}
    \lim_{\sigma\rightarrow 0}k(d^2(\us_i,\us_j);\sigma)=\delta(\us_i-\us_j)
    \label{eq:k-delta}
\end{equation}
where $\delta(\cdot)$ is the Dirac delta function. Furthermore, we assume that $d^2(\us_i,\us_j)$ is ``locally'' similar to the Euclidean distance:
\begin{equation}
    d^2(\us_i,\us_j) = (\us_i-\us_j)^\top \frac{1}{2}\left({\mathbf{H}(\us_i)+\mathbf{H}(\us_j)}\right)
            (\us_i-\us_j)+o(\|\us_i-\us_j\|^2),
\label{eq:d}
\end{equation}
when $\us_j$ is close to $\us_i$. In \req{eq:d}, $\mathbf{H}$ is a positive-definite matrix such that its inverse is $\mathbf{H}^{-1}=\uMH$ so that all GSTP formulas are comparable to those of MDM. We use the symmetric form $(\mathbf{H}(\us_i)+\mathbf{H}(\us_j))/2$ instead of $\mathbf{H}(\us_i)$ or $\mathbf{H}(\us_j)$ to preserve the symmetry of exchange $\us_i$
and $\us_j$.

The generalized kernel function $K_s$ is different from the Gaussian kernel $G_s$ (see \req{eq:feature_kernel}) in two ways. First, the Gaussian function is a special case of the generalized kernel, and thus other kernel functions that meet the above conditions can be used (e.g., $t$ distribution frequently used in manifold learning~\cite{maaten2008visualizing,maaten2009learning}). Second, the similarity function $d^2(\us_i,\us_j)$ does not need to be a distance metric, as we will demonstrate in \rsct{sec:ala}. It can be easily seen that the kernel function used in MDM is a special case of $K_s(\us_i,\us_j)$: 
\begin{equation}
    k(d^2(\us_i,\us_j);\sigma)=\exp\left(-\frac{1}{2\sigma^2}d^2(\us_i,\us_j)\right),
\end{equation}
where the similarity function is given as:
\begin{equation}
    d^2(\us_i,\us_j)=(\us_i-\us_j)^\top\frac{1}{2}(\uM^{-1}(\us_i)+\uM^{-1}(\us_j))(\us_i-\us_j).
\end{equation}

Similarly to MDM, the kernel function is not normalized in its general form due to position-dependent $\uMH(\us)$, where the distortion of the distribution can be calculated through a change of variable as $\rho_M(\us_i) \approx \detm{\uMH(\us_i)}^{1/2}\rho(\us_i)$ (see Supporting Information). The explicit expression of the transition matrix after row-normalization is:
\begin{equation}
P^\dub_{ij} = \frac{\frac{1}{\sqrt{\rho(\us_j)}} K_s(\us_i,\us_j) \detm{\uMH(\us_j)}^{-1/4}}
{\sum_k \frac{1}{\sqrt{\rho(\us_k)}} K_s(\us_i,\us_k)\detm{\uMH(\us_k)}^{-1/4}},
\label{eq:q-mij}
\end{equation}
where $\rho(\us)$ becomes the CV distribution (\req{eq:prob-s-equiv}). Note that \req{eq:q-mij} has been evaluated only with unbiased samples. 

Since $\uP^\dub$ is a transition matrix, we are interested in the invariant probability, $\pi$, of $\uP^\dub$, which satisfies $\sum_i \pi(\us_i) P^\dub_{ij}=\pi(\us_j)$. Solving $\pi(\us_i)$ requires calculating the left eigenvector of $\uP^\dub$ with the eigenvalue equal to one. We want to emphasize that $\pi(\us_i)$ is the invariant probability of the $i$'th state in the Markov chain of unbiased data instead of the Boltzmann distribution at $\us_i$.

We can prove that $\pi(\us_i)$ is a constant in the limit of $N\rightarrow\infty$ and $\sigma\rightarrow 0$. In DM or MDM, the fact that $\pi(\us_i)$ is a constant is consistent with the properties of $\mathcal{L}$. Our study demonstrates that $\pi(\us_i)$ is constant even if the sample Markov chain does not require a background diffusion process, as long as the kernel function properly approximates the Dirac delta function (see Supporting Information for a detailed proof).

\subsection{Case II: Biased Data}
\begin{figure}[ht]
    \centering
    \includegraphics{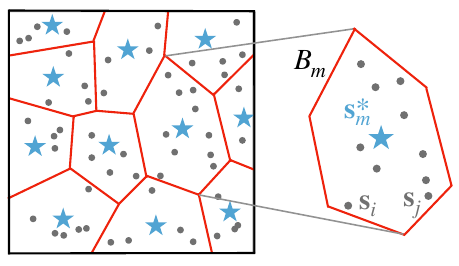}
    \caption{Unbiasing of the transition matrix implemented in GSTP. The biased samples $\pset{\sus_m}$ are partitioned in the CV space into cells that are macroscopic states (blue) that consist of the unbiased samples $\pset{\us_i}$ (grey). Each macroscopic state $B_m$ is defined around the corresponding biased sample $\sus_m$.}
    \label{fig:cg-illstrate}
\end{figure}

In order to derive the formula of $\uP$ with biased simulation data, we propose a hypothetical coarse-graining approach (\rfig{fig:cg-illstrate}). 
A collection of biased configurations, $\{\sus_i\}$ (blue dots), is spread throughout the CV space. Every configuration serves as the center of a Voronoi‑like region, so the full set partitions the space into discrete macrostates. The $m$'th macrostate, denoted $B_m$, is simply the region that envelops the configuration $\sus_m$. We want to emphasize that the coarse-graining scheme is used only in the theory development. Applying the derived formula to calculate $\uP^\db$ with biased simulation data does not require the coarse-graining method presented in this section. 

In our approach, we suggest that the GSTP from cell $m$ in step $t$ to cell $n$ in step $t+1$ can be viewed as the collection of all pairs of transition probabilities $T{_{ij}}$, such as $i \in B_m$ in step $t$ to $j\in B_n$ in step $t+1$:
\begin{align}
    P^\db_{mn} &= \operatorname{Pr}(\us(t+1)\in B_n \cond \us(t)\in B_m) \nonumber\\
    &= \frac{\sum\limits_{\us_i\in B_m,\us_j\in B_n} P^\dub_{ij}
        \pi(\us_i)}{\sum\limits_{\us_i\in B_m}\pi(\us_i)}
\label{eq:cg-Pmn}
\end{align}
where $P^\dub_{ij}=\operatorname{Pr}(\us(t+1)=\us_j \cond \us(t)=\us_i)$ defined in \rsct{sec:gstp-unbiased}. We will provide a detailed explanation of this in the following Section. Considering that $\pi(\us_i)$ is a constant, \req{eq:cg-Pmn} becomes:
\begin{equation}
P^\db_{mn} \approx \frac{1}{N_m} \sum_{\us_i\in B_m,\us_j\in B_n}P^\dub_{ij}
\label{eq:qpmn}
\end{equation}
where $N_m$ is the number of unbiased samples in $B_m$. 

If the biased dataset is sufficiently large, the macrostate $B_m$ is small enough such that the probability density $\rho(\us)$ is approximately constant within $B_m$, which allows us to assume that $\rho(\us_i)\approx\rho(\sus_m)$. Next, if each Voronoi cell is much smaller than the broadening of $K_s$, then $\uMH(\us_i)\approx\uMH(\sus_m)$. Therefore, the kernel $K_s(\us_i,\us_j)\approx K_s(\sus_m,\sus_n)$. Taking all this together, \req{eq:qpmn} becomes:
\begin{equation}
P_{mn} \approx 
\frac{\frac{N_n}{\sqrt{\rho(\sus_n)}} K_s(\sus_m,\sus_n)\detm{\uM(\sus_n)}^{-1/4}}
{\sum_l \frac{N_l}{\sqrt{\rho(\sus_l)}} K_s(\sus_m,\sus_l)\detm{\uM(\sus_l)}^{-1/4}},
\label{eq:Pmn_N}
\end{equation}
where the number of samples in the $n$'th state can be approximated as $N_n\approx \rho(\sus_n) V_n$, where $V_n\approx 1/\tilde{\rho}(\sus_n)$ is the volume of $B_n$ and $\tilde{\rho}(\sus_n)$ is the biased probability density evaluated at $\sus_n$. Consequently, the number of samples is $N_n\approx \rho(\tus_n)/\tilde{\rho}(\tus_n)=\omega_n$, where $\omega_n$ is the unbiasing weight. Finally, \req{eq:Pmn_N} can be rewritten as:
\begin{equation}
P_{mn} \approx \frac{\frac{\omega_n}{\sqrt{\rho(\sus_n)}} K_s(\sus_m,\sus_n)
\detm{\uMH(\sus_n)}^{-1/4}}
{\sum_l \frac{\omega_l}{\sqrt{\rho(\sus_l)}} K_s(\sus_m,\sus_l)\detm{\uMH(\sus_l)}^{-1/4}},
\label{eq:final}
\end{equation}
which is the main result of this work. For a detailed derivation, see Supporting Information. We want to emphasize that the coarse-graining process described in Fig.\ref{fig:cg-illstrate} is not required when using \req{eq:final}.

Equation~\ref{eq:final} is equivalent to the transition matrix in MDM constructed from biased samples when the kernel function is given by a heterogeneous Gaussian in the feature space (\req{eq:feature_kernel}). In deriving \req{eq:hete_gauss}, it is necessary to assume that the transition matrix $T_{ij}$ converges to the infinitesimal generator of the diffusion process for unbiased simulation. However, our derivation demonstrates that this assumption applies only to a specific case within the broader family of kernel functions that GSTP can use. Therefore, GSTP provides a straightforward and general approach that can accommodate various kernel functions without requiring the generalized transition matrix $P_{mn}$ to converge to the generator of any diffusion process.

\section{Results and Discussion}
In this Section, we will present and discuss several numerical examples to verify our main result (\req{eq:final}). We will generate $P_{mn}$ with biased and unbiased data using various kernel functions $K_s$. We will start with a one-dimensional harmonic oscillator for which the exact solution of diagonalizing the infinitesimal generator is known. The proposed methods will be further validated by examples of alanine dipeptide in vacuum and met-enkephalin in water. The details of MD simulations are described in Supporting Information.

\subsection{Kernel Functions in the Coordinate Space} 

In the following, we will verify our derivation of the transition probability $P_{mn}$ using kernels built in the coordinate space $\ux$. In such cases, GSTP reduces to DM where $\detm{\uMH}^{-1/4}$ in \req{eq:final} is a constant. Therefore, the main result simplifies to:
\begin{equation}
P_{mn} \approx \frac{\frac{\omega_n}{\sqrt{\rho(\xus_n)}} K(\xus_m,\xus_n)}{\sum_l \frac{\omega_l}{\sqrt{\rho(\xus_l)}} K(\xus_m,\xus_l)},
\label{eq:cg-tm-final}
\end{equation}
where, as before, we denote biased samples by $\xus$, and $\omega$ are the corresponding weights.

\begin{figure*}[p]
    \centering
    \includegraphics[width=1.0\textwidth]{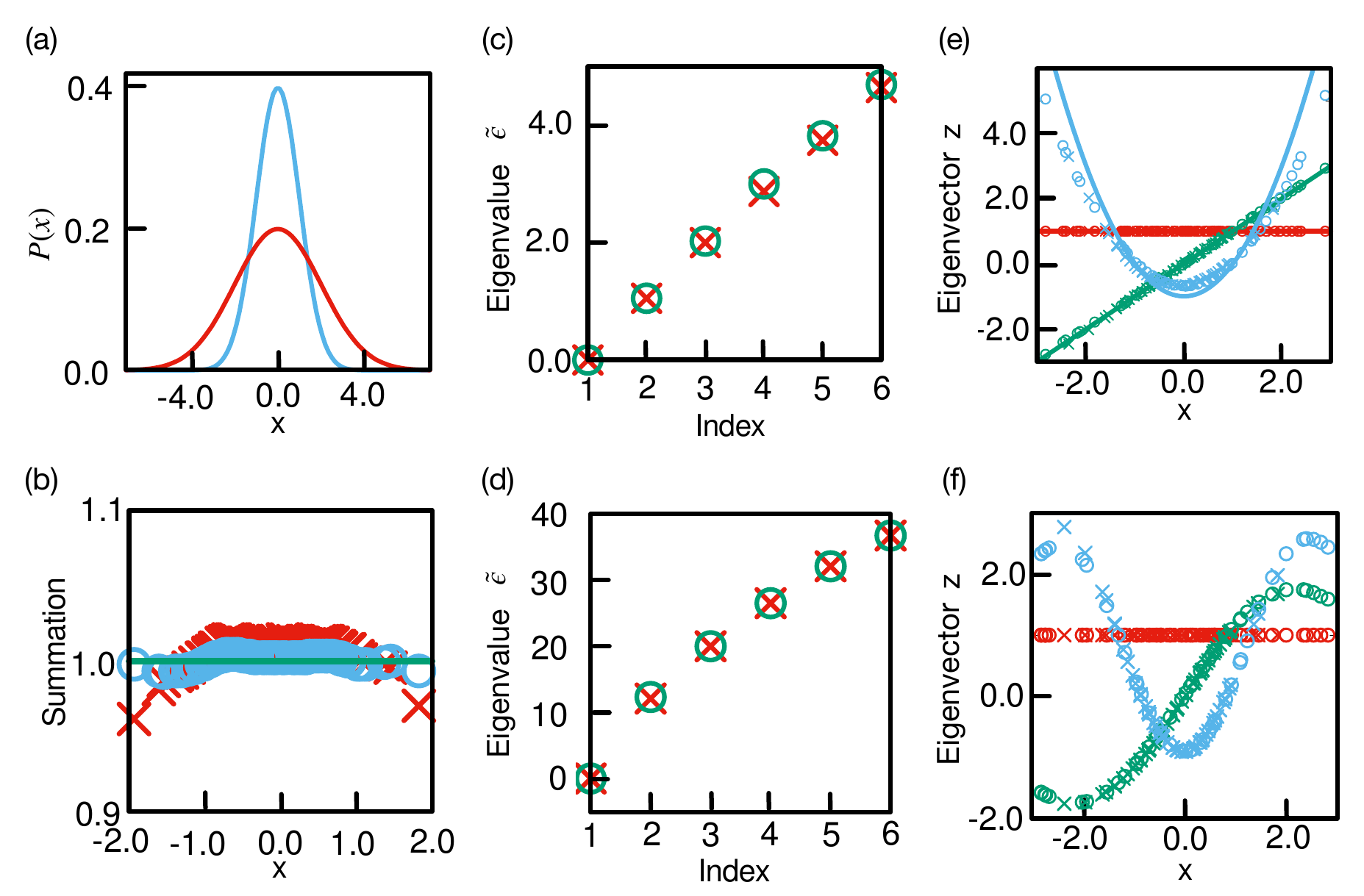}
    \caption{Validation of results for the 1D harmonic oscillator. (a) The Boltzmann distribution of $U(x)=\frac{1}{2}x^2$ at $\beta=1$ (blue) and the biased distribution of the same potential energy at $\beta=0.5$ (red). (b) The comparison of the row-summation for transition matrices built from the unbiased data using the Gaussian kernel (blue) and the $t$ kernel (red). (c-d) The scaled eigenvalues $\tilde{\epsilon}$ from the unbiased data (green) and the biased data (red). The Gaussian kernel was used in (c), while the $t$ kernel was used in (d). (e-f) The uniformly selected elements from the normalized eigenvector $z$ as a function of $x$ with the Gaussian kernel (e) and the $t$ kernel (f). The smallest, second smallest, and third smallest scaled eigenvalues $\tilde{\epsilon}$ in red, green, and blue. The cross and circle markers correspond to $z$ from the unbiased and biased data, correspondingly.}
    \label{fig:1D}
\end{figure*}

\subsubsection{Validation: 1D Harmonic Oscillator}
We validate GSTP using a simple example: a one-dimensional harmonic oscillator with potential energy functions $U(x) = \frac{1}{2} x^2$. We further assume that the dynamics is given by a diffusion process (\req{eq:ger_co}). The $n$'th eigenvalue of the diffusion process generator is $\epsilon_n = n$ for a non-negative integer $n$, and the $n$'th eigenvector $\psi_n(x)$ is the $n$'th order Hermite polynomial. We construct two MDM transition matrices, the first from unbiased data $\uT^{\rm{ub}}$ at $\beta=1$ and the second from biased data sampled at $\beta=0.5$, $\uT^{\rm{b}}$ (see \rfig{fig:1D}a). Details on constructing both MDMs can be found in Supporting Information. We will further replace the Gaussian kernel in MDM with a student t-distribution so that $\uP^{\rm{ub}}$ will be generated with the unbiased data and $\uP^{\rm{b}}$ with the biased data. Details on the construction of MDMs and GSTPs can be found in Supporting Information.

As shown in \rfig{fig:1D}, biased data (red) and unbiased data (blue) have different distributions. Our aim is to obtain the transition probabilities ($m$ and $n$) with the biased data distribution (set as $T^{\rm b}_{mn}$ or $P^{\rm b}_{mn}$), which cannot be obtained by constructing a GSTP directly from biased data. Constructing GSTP directly follows \req{eq:final}. 
 
We first test whether the row-sum $\sum_i T^{\rm{ub}}_{ij}$ is approximately equal to one, as discussed in \rsct{sec:gstp-unbiased}. As illustrated by the blue circles in \rfig{fig:1D}b, $\sum_i T^{\rm{ub}}_{ij}$ fluctuate tightly around 1, indicating that the estimate is accurate. That $\uT^{\rm{ub}}$ and $\uT^{\rm{b}}$ contain the same information can be shown by comparing the eigenvalues and eigenvectors of both MDM transition matrices. Recall that we denote the scaled eigenvalue and normalized eigenvectors by $\tilde{\epsilon}_n\equiv 2(\lambda_n-1)/\sigma^2$ and $z_n$, respectively. Following the discussions in \rsct{sec:dms}, we know that $\epsilon_n\approx\tilde{\epsilon}_n$ and the $i$'th element of $z_n$ is approximately $\psi_n(x_i)$. \rfig{fig:1D}c shows that the eigenvalues $\tilde{\epsilon}$ from both $\uT^{\rm{ub}}$ and $\uT^{\rm{b}}$ agree with our theoretical derivation for $\epsilon$. Similarly, \rfig{fig:1D}d suggests that the normalized eigenvectors $z$ of the unbiased and biased data are consistent and agree with the derived expression for $\psi(x)$ up to normalization (see Supporting Information for additional details on normalizing the eigenvectors).

Using the same datasets, we check that \req{eq:final} remains valid with a non-Gaussian kernel. For this, we use a $t$ kernel function:
\begin{equation}
     K_{st}(x_i,x_j) = {\left(1 + \frac{1}{2\sigma^2} \|x_i-x_j\|^2\right)^{-\frac{W + 1}{2}}},
\end{equation}
where $W$ is the parameter that controls the heaviness of its tails and $\sigma$ controls the broadening. We set $W$ to 1 so that $\lim_{\sigma\rightarrow 0}K_{st}(x,x')=\delta(x-x')$. Similarly to the Gaussian kernel, we test whether $\sum_iP^{\rm ub}_{ij}\approx 1$; see \rfig{fig:1D}b (red). Although the conservation of $\sum_i P^{\rm ub}_{ij}$ is not as close to 1 as $\sum_i T^{\rm{ub}}_{ij}$, the error is still within $\approx$3\% in a range with sufficient data support. We notice that $\sum_i P^{\rm ub}_{ij}$ deviates from 1 systematically when $\pnrm{x_i-x_j}$ is large, probably due to the slow decay of the $t$ kernel function (see Supporting Information for a discussion). However, the conservation of $\sum_i P^{\rm ub}_{ij}$ is good enough to guarantee the validation of \req{eq:final}, as shown in \rfig{fig:1D}d,e. Due to the use of a non-Gaussian kernel, the scaled eigenvalues and eigenvectors of $\mathbf{P}^{\rm ub}$ and $\mathbf{P}^{\rm b}$ are different from $\epsilon$ and $\psi(x)$. This is because $\uP$ may no longer asymptotically converge to the generator of \req{eq:qsde}. However, the eigenvalues and eigenvectors of $\uT_{st}$ and $\mathbf{P}_{st}$ are consistent. It suggests that \req{eq:final} is capable of generating a consistent GSTP with unbiased and biased data, even if such GSTP is not explicitly related to a Langevin equation. 

\subsubsection{Validation: Alanine Dipeptide in Vacuum}
\label{sec:ala_vac}
\begin{figure}[H]
    \centering
    \includegraphics[width=1.0\textwidth]{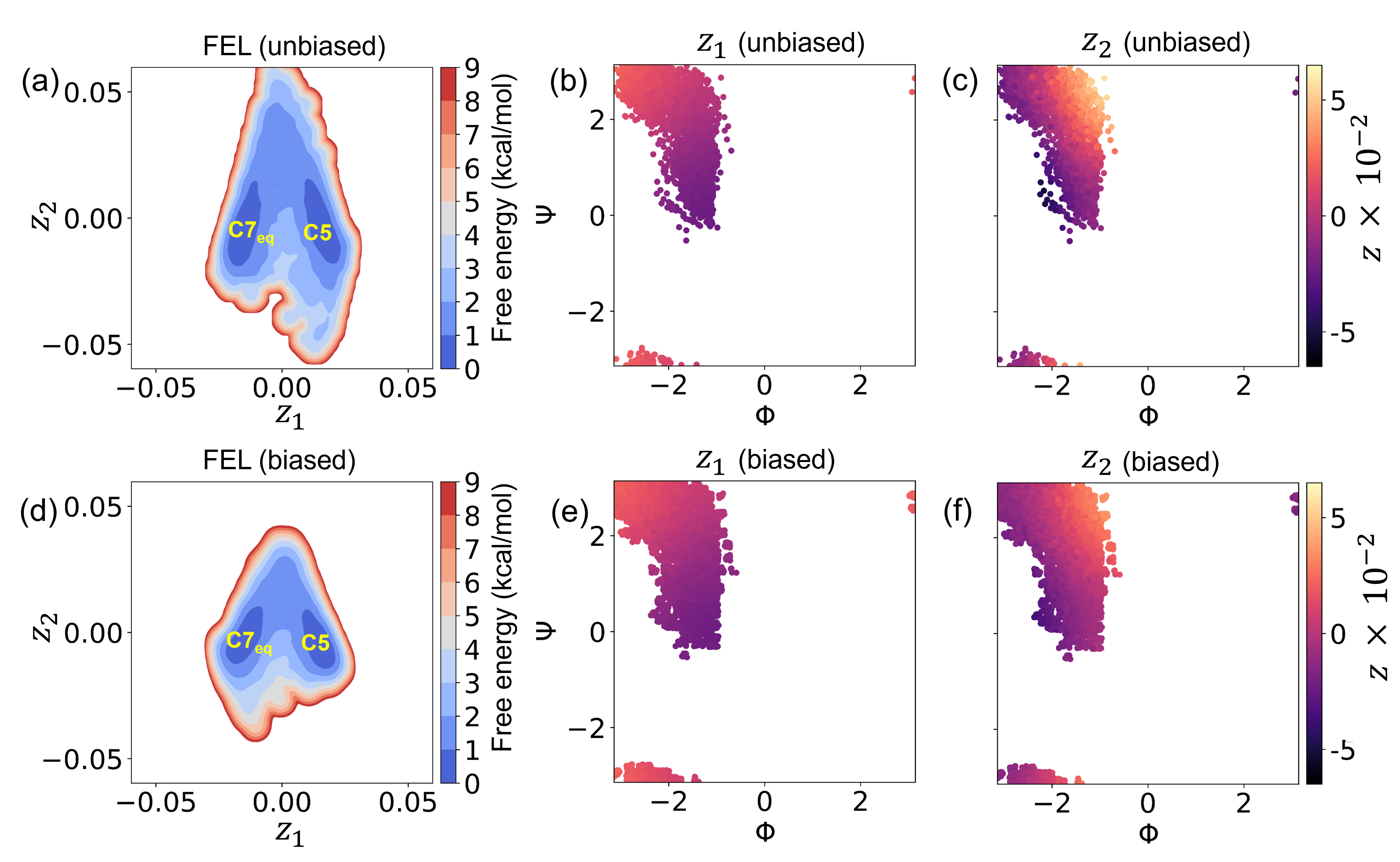}
    \caption{
    GSTP constructed from unbiased MD simulation and biased enhanced sampling data. The kernel function defined in \req{eq:kernel_weights} is used. The two slowest modes, $z_1$ and $z_2$, are calculated. In (a), we use data from the plain MD simulation to construct the FEL, and in (d), we employ unbiased data from the enhanced sampling simulation to construct the FEL. Note that conformations that are sampled in the enhanced sampling simulation and not in the plain MD simulation are removed to ensure direct comparison. Panels (b,c) ((e,f)) map \( \text{z}_1 \) (b) (\( \text{z}_1 \) (e)) and \( \text{z}_2 \) (c) (\( \text{z}_2 \) (f)) to the backbone torsion space. Color brightness indicates the $z$ values from negative (dark) to positive (bright). 
    }
    \label{fig:6}
\end{figure}

Next, we will assess the accuracy of \req{eq:cg-tm-final} by using enhanced sampling data obtained from simulations of alanine dipeptide in vacuum. For this, we employ well-tempered metadynamics (WTM) to drive the sampling, using the Ramachandran torsion angles $\Phi$ and $\Psi$ as variables to enhance fluctuations. Another set of samples is generated with a plain MD simulation to reveal the equilibrium distribution. As can be seen, the unbiased simulation cannot efficiently sample the energy minima as they are separated by a high barrier (see \rfig{fig:6}). For comparison, we analyze the states with which the conformational transitions are sampled sufficiently in the plain MD, that is, C5 and C7$_{eq}$ (see \rfig{fig:6}). Details of MD simulations and WTM are listed in Supporting Information. 

To show that GSTP can effectively employ various kernel functions, we consider a kernel where the distance between the atomic coordinate vectors (after RMSD alignment) is used as a similarity function:
\begin{equation}
K(\ux_i,\ux_j)=\frac{1}{(2\pi\sigma^2)^{\frac{D}{2}}}
\exp\left(-\frac{\|\alpha\circ(\ux_i-\ux_j)\|^2}{2\sigma^2}\right)\;\;\ldotp
\label{eq:kernel_weights}
\end{equation}
where $\alpha$ is a weighting vector. The backbone atoms were assigned a weight ten times greater than those of other non-hydrogen atoms to highlight their significance in the description of the slow modes of alanine dipeptide.

We use the kernel function $K$ to build GSTP with \req{eq:cg-tm-final}. The weight $\omega$ for each bin on the GSTP graph can be accessed from the reweighing of the enhanced samples. We use the Tiwary-Parrinello approach~\cite{tiwary2015time}. Other methods can also be used for unbiasing~\cite{giberti2020iterative}.

To demonstrate the effectiveness of GSTP with enhanced sampling data, we calculated the first two eigenvectors, $z_1$ and $z_2$, of GSTP with plain MD samples. We also evaluated $z_1$ and $z_2$ of GSTP with enhanced sampling data. They represent the two slowest modes of the system. The eigenvectors in \rfig{fig:6}a are from GSTP using a plain MD simulation, and the eigenvectors in \rfig{fig:6}d are from GSTP with \req{eq:cg-tm-final} and enhanced sampling data. Note that only basins appearing in the plain MD simulation are kept in the case of biased enhanced sampling to ensure direct comparison. As shown in \rfig{fig:6}a,d, the structure of GSTP with enhanced sampling samples and GSTP from plain MD samples are close in the vector space of the first two diffusion coordinates. In addition, a consistent distribution of $z_1$ and $z_2$ in the vector space of $\Phi$ and $\Psi$ is observed in both cases (\rfig{fig:6}b,c and e,f).

Our results validate that the application of \req{eq:cg-tm-final} to approximate GSTP with customized kernel functions yields consistent results between the plain MD simulation and the biased enhanced sampling simulation, as long as the convergence conditions are satisfied.

\subsection{Kernel Functions in the Feature Space}
In the following validation, we will present the validation of \req{eq:final} for GSTP in a feature space with two examples: alanine dipeptide in vacuum and met-enkaphalin peptide in aqueous solution. Unlike the last Section, in which atomistic Cartesian coordinates are used to describe structural similarity in the kernel function, the kernel functions for GSTP are defined on feature similarity. Specifically, torsion angles are selected as features for GSTP. 

\subsubsection{Validation: Alanine Dipeptide in Vacuum}
\label{sec:ala}

We consider the kernel function in the feature space $\us=\uq(\ux)$ in order to apply \req{eq:final}. Torsion angles are widely used as CVs as they often represent slow motions during conformation changes. To use the torsion angles $\pmb{\theta} = \ppar{\theta_1, \dots, \theta_n}$ as the features for \req{eq:feature_kernel}, and account for the periodicity, we calculate the geometric difference between pairs of samples with periodic boundary conditions in [$0, 2\pi]$ using the minimum image convention:
\begin{equation}
    \theta_l^{ij} = \theta_l^i - \theta_l^j-2\pi\pbkt*{\frac{\theta_l^i - \theta_l^j}{2\pi}}
\end{equation}
where $\theta_l^i$ denotes the $l$'th torsion angle of the $i$'th sample in the dataset and $\pbkt*{x}$ is the nearest-integer function. We denote $\pmb{\theta}^{ij}=(\theta_1^{ij},\theta_2^{ij},\cdots,\theta_n^{ij})^\top$.

We first test the following kernel:
\begin{equation}
K_{\pmb{\theta}}(\pmb{\theta}_{i},\pmb{\theta}_j) = \exp\ppar*{-\frac{1}{4\sigma^2}{\pmb{\theta}^{ij}}^\top(\uM^{-1}(\pmb{\theta}_i)+ 
\uM^{-1}(\pmb{\theta}_j))\pmb{\theta}^{ij}}\ldotp
\label{eq:kernel_theta}
\end{equation}
\begin{figure}[H]
    \centering
    \includegraphics[width=1.0\textwidth]{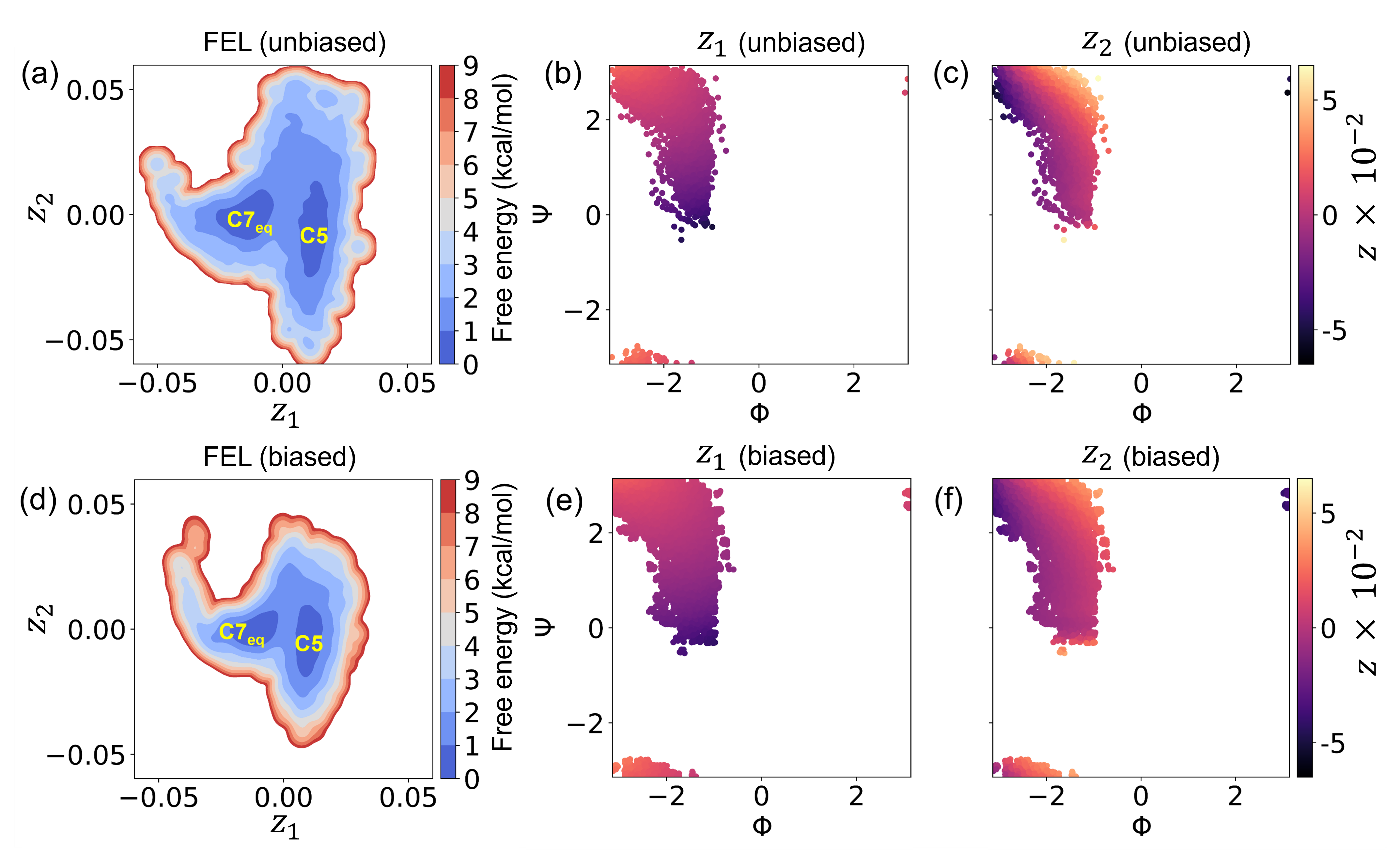}
    \caption{GSTP constructed from unbiased MD simulation and biased enhanced sampling data. The kernel function defined in \req{eq:kernel_theta} was used to estimate the two slowest motions, $z_1$ and $z_2$. Similar to \rfig{fig:6}, Panel (a) uses data from the plain MD simulation to construct the FEL, and panel (d) uses the unbiased data from the enhanced sampling simulation to construct the FEL. Panels (b,c) ((e,f)) map \( \text{z}_1 \) (b) (\( \text{z}_1 \) (e)) and \( \text{z}_2 \) (c) (\( \text{z}_2 \) (f)) to the backbone torsion space. The color brightness indicates the eigenvector values change from negative (dark) to positive (bright).}
    \label{fig:7}
\end{figure}

As shown in \rfig{fig:7}a,d, GSTP constructed from biased samples closely matches GSTP from unbiased samples in the space of $z_1$ and $z_2$. Furthermore, a consistent distribution of $z_1$ and $z_2$ in the vector space of $\Phi$ and $\Psi$ is observed in both cases (\rfig{fig:7}b,c and e,f). Note that the resulting eigenvectors slightly differ from those presented in \rfig{fig:6}, implying that the eigenvectors of GSTP depend on the used kernel function. Nonetheless, the separation between the basins is evident. These findings validate \req{eq:final} with enhanced sampling data and a kernel function in the feature space.

To demonstrate that \req{eq:final} can be used for various kernel functions defined in the feature space, we construct a new kernel function by introducing a torsion angle similarity: $v_l^{ij} \equiv \sin(\theta_l^{ij}/2)$, which maps $\theta_l^{ij}\in (-\pi,\pi]$ to $v_l^{ij}\in (-1,1]$. By defining $\uv^{ij}=(v_1^{ij},\cdots,v_n^{ij})^\top$, the kernel function becomes:
\begin{align}
    K_{\uv}(\pmb{\theta}_i,\pmb{\theta}_j) = \exp\ppar*{-\frac{1}{\sigma^2}{\uv^{ij}}^\top (\uM^{-1}(\pmb{\theta}_i)+\uM^{-1}(\pmb{\theta}_j)) {\uv^{ij}}}.
\label{eq:feature_kernel_sin}
\end{align}

\begin{figure}[H]
    \centering
    \includegraphics[width=1.0\textwidth]{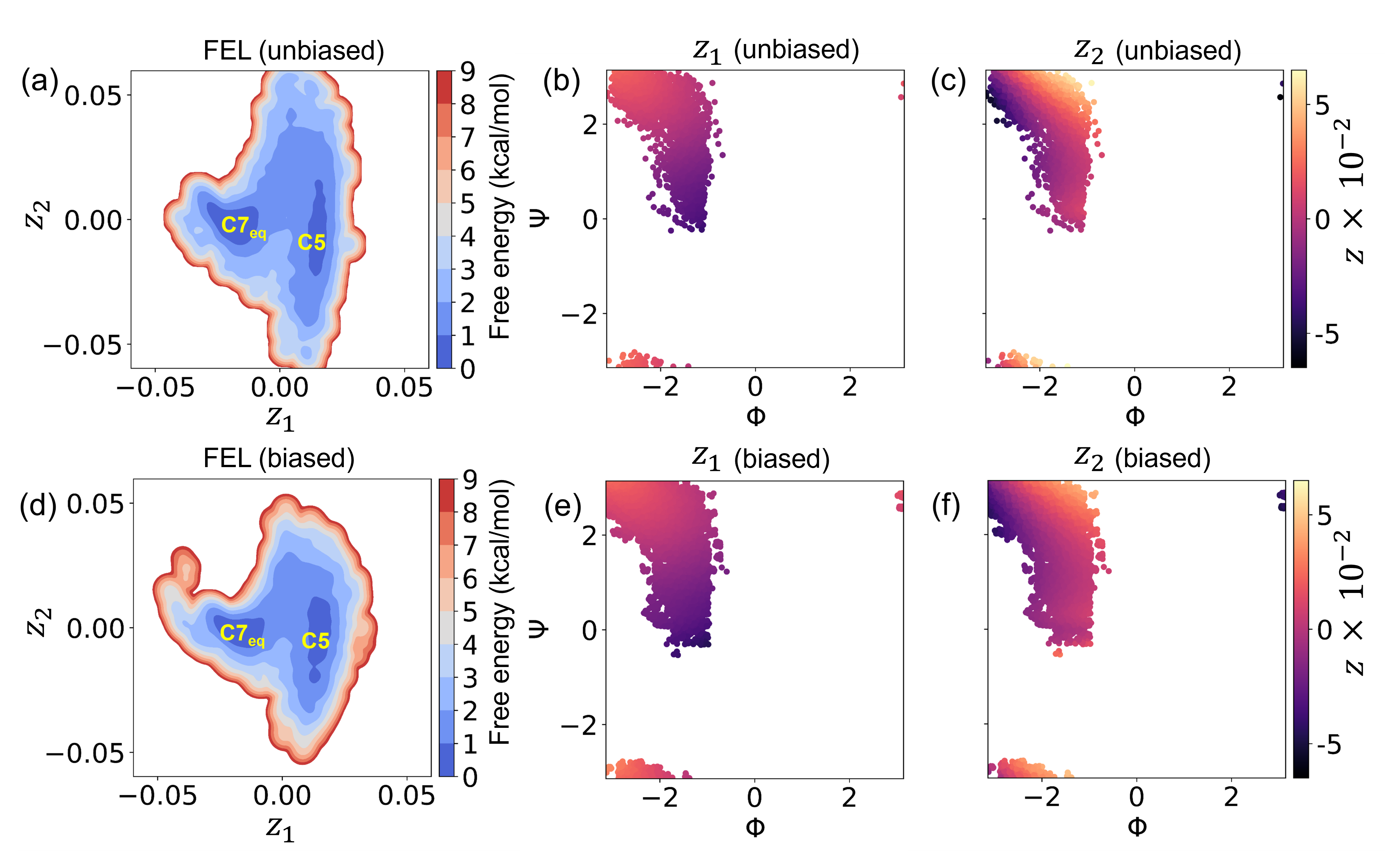}
    \caption{GSTP constructed from unbiased MD simulation and biased enhanced sampling data. The kernel function defined in \req{eq:feature_kernel_sin} was used to construct the two slowest motions, denoted as $z_1$ and $z_2$. Panel (a) shows data from the plain MD simulation to construct the FEL, and panel (d) presents unbiased data from the enhanced sampling simulation to construct the FEL. Panels (b,c) ((e,f)) map \( \text{z}_1 \) (b) (\( \text{z}_1 \) (e)) and \( \text{z}_2 \) (c) (\( \text{z}_2 \) (f)) to the backbone torsion space. The color brightness indicates the eigenvector values change from negative (dark) to positive (bright).}
    \label{fig:9}
\end{figure}

\rfig{fig:9} demonstrates that \req{eq:final} generates consistent GSTPs by using unbiased MD or enhanced sampling datasets, even if the ``feature similarity'' is highly non-linear. Furthermore, the FELs in the space of eigenvectors constructed by GSTP with different kernels are still similar despite using different kernels. This suggests that the GSTP algorithm is robust with various kernel function choices.

\subsubsection{Validation: Met-enkephalin in Water}
In this Section, we benchmark \req{eq:final} with a pentapeptide named Met-enkephalin in water, which is a standard benchmarking example in many computational studies~\cite{sanbonmatsu2002structure,evans2003free}. Since exploring all conformations for this system is challenging and requires long plain MD simulations, we compared the GSTP results generated from two different enhanced sampling methods. The first is WTM, while the second is TAMD-driven adiabatic free-energy dynamics (TAMD/d-AFED)~\cite{abrams2008efficient,rosso2002use,maragliano2006temperature}. To drive the sampling in WTM, we used stochastic kinetic embedding (StKE), a machine learning technique to learn CVs~\cite{zhang2018unfolding}. For TAMD/d-AFED, we used ten Ramachandran torsion angles. Both these enhanced sampling simulations, including the construction of StKE CVs and the estimation of unbiasing weights, were generated in our previous study~\cite{liu2024unbiasing}.

We tested \req{eq:final} by constructing the GSTP in a feature space spanned by two Ramachandran torsion angles, ($\Phi$ and $\Psi$) of the fourth residue of met-enkephalin. The reason for selecting only one pair of
Ramachandran torsion angles is to reduce the costs of constructing the diffusion tensor $\uM$. We used the kernel function given by \req{eq:kernel_theta} in this example. 

\begin{figure}[H]
    \centering
    \includegraphics[width=1.0\textwidth]{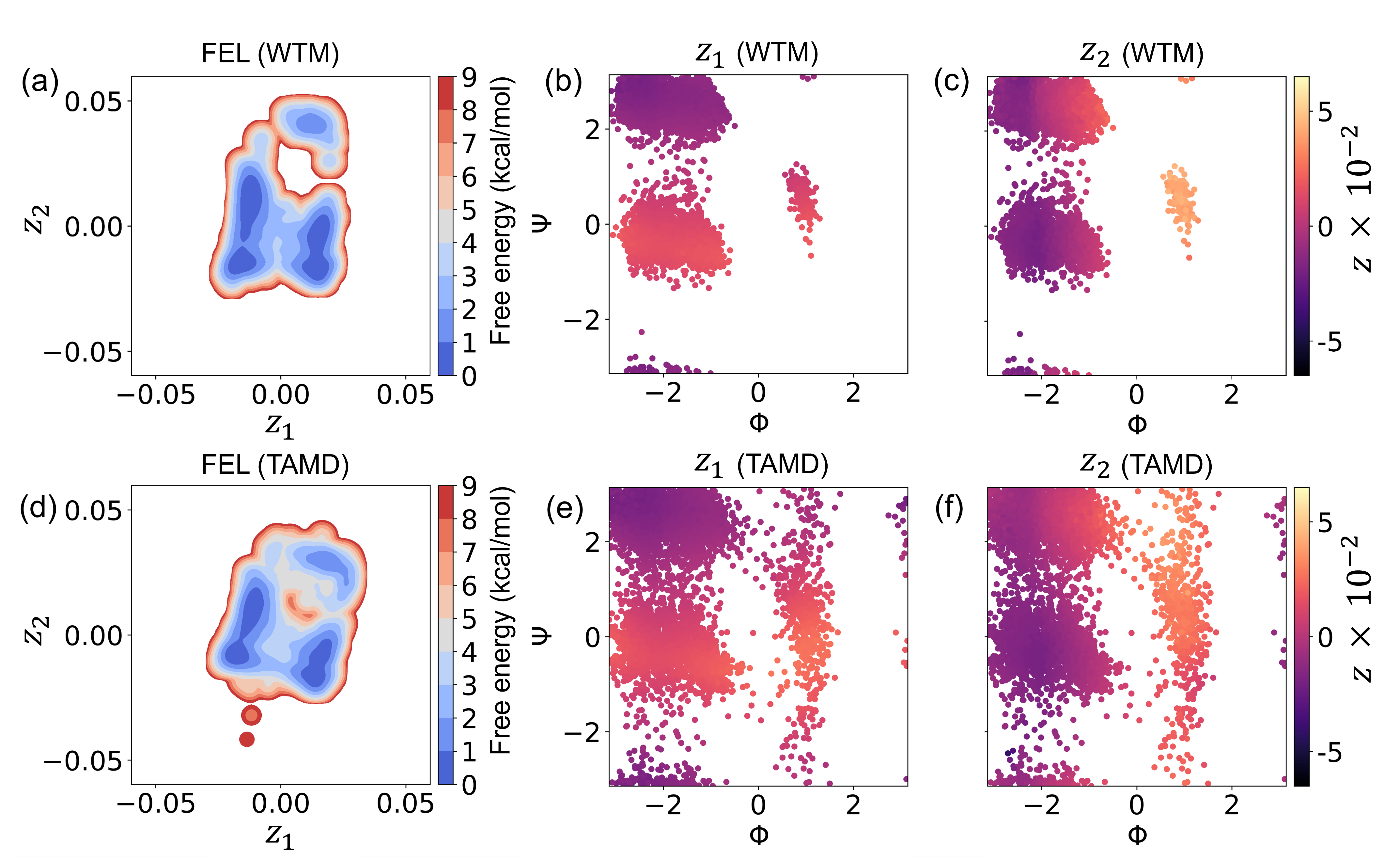}
    \caption{GSTP constructed from two different biased enhanced sampling methods: WTM and TAMD. The kernel function defined in \req{eq:kernel_theta} was used using $\Phi$ and $\Psi$ of the fourth residue of met-enkephalin. The two slowest modes, $z_1$ and $z_2$, are calculated. Panel (a) projects structures from the WTM simulation to construct the FEL, and panel (d) projects structures from the TAMD simulation to construct the FEL. Panels (b,c) ((e,f)) map \( \text{z}_1 \) (b) (\( \text{z}_1 \) (e)) and \( \text{z}_2 \) (c) (\( \text{z}_2 \) (f)) to the backbone torsion space. The color brightness indicates the eigenvector values change from negative (dark) to positive (bright).}
    \label{fig:8}
\end{figure}

As in \rsct{sec:GSTP}, we include the datasets generated by both simulations. GSTPs computed from WTM and TAMD/d-AFED samples are consistent in the space of $z_1$ and $z_2$. Both FELs in \rfig{fig:8}a,d show five minima with similar free energy differences, and the barrier heights connecting different free energy minima are similar when comparing two FELs. It demonstrates that using \req{eq:final} is not sensitive to the enhanced sampling methods used in the simulation, as long as these methods generate the Boltzmann distribution after unbiasing.

\section{Conclusions}
In this study, we present a technique for constructing generalized sample transition probability (GSTP) from biased enhanced sampling simulations, broadening the ability to derive slow motions from biased enhanced sampling simulations. Unlike the standard diffusion map (DM) or the Mahalanobis diffusion map (MDM), which directly relies on the underlying stochastic process, our technique uses a coarse-graining procedure in the coordinate space or the feature space to derive the unbiasing formula for GSTP. By decoupling GSTP and stochastic process, the unbiasing formula can be applied to non-Gaussian kernel functions or various molecular structural similarity metrics, other than those used in DM and MDM. By constructing a generalized transition matrix with additional reweighting, GSTP approximates DM and MDM, which requires unbiased data that are often inaccessible.

Our approach is adaptable, demonstrating its capability to accurately recover kinetic information from diverse systems and kernel functions, including those not originally derived based on underlying stochastic processes. We have validated the robustness of GSTP building transition matrices from various sampling strategies, confirming its independence from specific enhanced sampling methods. This adaptability ensures that GSTP can be widely applied.

Importantly, GSTP offers a straightforward extension to recent machine learning techniques. For example, it can be combined with spectral map~\cite{rydzewski2023spectral,rydzewski2024learning,rydzewski2024tse} to achieve higher accuracy in maximizing timescale separation to construct slow CVs and their corresponding FELs from biased simulations. GSTP can also be integrated with a feature selection pipeline built on top of DMs~\cite{rydzewski2023selecting}. Techniques such as StKE~\cite{zhang2018unfolding} or multiscale reweighted stochastic embedding~\cite{rydzewski2021multiscale,rydzewski2022reweighted}, among many others that also incorporate the estimation of pairwise transition probabilities~\cite{chen2021collective,rydzewski2023manifold}, can profit from our method.

In summary, the GSTP framework provides a general machine learning framework to derive kinetic information solely based on thermodynamics, represented by datasets from biased simulations. The implemented technique ensures the design and optimization of slow CVs with greater flexibility and accuracy. We believe that GSTP will be applicable to more complex processes of physical and biological importance.

\begin{acknowledgement}
J. R. acknowledges funding from the Ministry of Science and Higher Education in Poland and the National Science Center in Poland (Sonata 2021/43/D/ST4/00920, ``Statistical Learning of Slow Collective Variables from Atomistic Simulations''). Y.W. and M.C. acknowledge the support of the American Chemical Society Petroleum Research Fund (Grant Number 67307).
\end{acknowledgement}

\begin{suppinfo}
Supporting Information is available free of charge at \url{https://pubs.acs.org/}.
\begin{itemize}
  \item Simulation details, including alanine dipeptide, met-enkephalin, and the estimation of the diffusion matrix.
  \item Kernel normalization in the feature space.
  \item Detailed derivation of \req{eq:hete_gauss} with the infinitesimal generator.
  \item Detailed derivation of transition probability coarse-graining.
  \item Additional details of the 1D problem, including unbiasing distribution of biased data and the normalization of eigenvectors.
  \item Code and data: Code is available at \href{https://github.com/wang5115/generalized-sample-transition-probability-GSTP-/tree/main}{GitHub}, and data at \href{https://drive.google.com/drive/folders/1DdJsEmw_vxVUWROrb_bTCR6YCgZkgLzf?usp=drive_link}{Google Drive}.

\end{itemize}
\end{suppinfo}

\bibliography{unbias_kinetics}

\providecommand{\latin}[1]{#1}
\makeatletter
\providecommand{\doi}
  {\begingroup\let\do\@makeother\dospecials
  \catcode`\{=1 \catcode`\}=2 \doi@aux}
\providecommand{\doi@aux}[1]{\endgroup\texttt{#1}}
\makeatother
\providecommand*\mcitethebibliography{\thebibliography}
\csname @ifundefined\endcsname{endmcitethebibliography}  {\let\endmcitethebibliography\endthebibliography}{}
\begin{mcitethebibliography}{82}
\providecommand*\natexlab[1]{#1}
\providecommand*\mciteSetBstSublistMode[1]{}
\providecommand*\mciteSetBstMaxWidthForm[2]{}
\providecommand*\mciteBstWouldAddEndPuncttrue
  {\def\EndOfBibitem{\unskip.}}
\providecommand*\mciteBstWouldAddEndPunctfalse
  {\let\EndOfBibitem\relax}
\providecommand*\mciteSetBstMidEndSepPunct[3]{}
\providecommand*\mciteSetBstSublistLabelBeginEnd[3]{}
\providecommand*\EndOfBibitem{}
\mciteSetBstSublistMode{f}
\mciteSetBstMaxWidthForm{subitem}{(\alph{mcitesubitemcount})}
\mciteSetBstSublistLabelBeginEnd
  {\mcitemaxwidthsubitemform\space}
  {\relax}
  {\relax}

\bibitem[Lindorff-Larsen \latin{et~al.}(2011)Lindorff-Larsen, Piana, Dror, and Shaw]{lindorff2011fast}
Lindorff-Larsen,~K.; Piana,~S.; Dror,~R.~O.; Shaw,~D.~E. {How Fast-Folding Proteins Fold}. \emph{Science} \textbf{2011}, \emph{334}, 517--520\relax
\mciteBstWouldAddEndPuncttrue
\mciteSetBstMidEndSepPunct{\mcitedefaultmidpunct}
{\mcitedefaultendpunct}{\mcitedefaultseppunct}\relax
\EndOfBibitem
\bibitem[Neha \latin{et~al.}(2023)Neha, Tiwari, Mondal, Kumari, and Karmakar]{neha2022collective}
Neha; Tiwari,~V.; Mondal,~S.; Kumari,~N.; Karmakar,~T. {Collective Variables for Crystallization Simulations--from Early Developments to Recent Advances}. \emph{ACS Omega} \textbf{2023}, \emph{8}, 127--146\relax
\mciteBstWouldAddEndPuncttrue
\mciteSetBstMidEndSepPunct{\mcitedefaultmidpunct}
{\mcitedefaultendpunct}{\mcitedefaultseppunct}\relax
\EndOfBibitem
\bibitem[Beyerle \latin{et~al.}(2023)Beyerle, Zou, and Tiwary]{beyerle2023recent}
Beyerle,~E.~R.; Zou,~Z.; Tiwary,~P. {Recent Advances in Describing and Driving Crystal Nucleation using Machine Learning and Artificial Intelligence}. \emph{Curr. Opin. Solid State Mater. Sci.} \textbf{2023}, \emph{27}, 101093\relax
\mciteBstWouldAddEndPuncttrue
\mciteSetBstMidEndSepPunct{\mcitedefaultmidpunct}
{\mcitedefaultendpunct}{\mcitedefaultseppunct}\relax
\EndOfBibitem
\bibitem[Piccini \latin{et~al.}(2022)Piccini, Lee, Yuk, Zhang, Collinge, Kollias, Nguyen, Glezakou, and Rousseau]{piccini2022ab}
Piccini,~G.; Lee,~M.-S.; Yuk,~S.~F.; Zhang,~D.; Collinge,~G.; Kollias,~L.; Nguyen,~M.-T.; Glezakou,~V.-A.; Rousseau,~R. {Ab Initio Molecular Dynamics with Enhanced Sampling in Heterogeneous Catalysis}. \emph{Catal. Sci. Technol.} \textbf{2022}, \emph{12}, 12--37\relax
\mciteBstWouldAddEndPuncttrue
\mciteSetBstMidEndSepPunct{\mcitedefaultmidpunct}
{\mcitedefaultendpunct}{\mcitedefaultseppunct}\relax
\EndOfBibitem
\bibitem[Baron and McCammon(2013)Baron, and McCammon]{baron2013molecular}
Baron,~R.; McCammon,~J.~A. {Molecular Recognition and Ligand Association}. \emph{Annu. Rev. Phys. Chem.} \textbf{2013}, \emph{64}, 151--175\relax
\mciteBstWouldAddEndPuncttrue
\mciteSetBstMidEndSepPunct{\mcitedefaultmidpunct}
{\mcitedefaultendpunct}{\mcitedefaultseppunct}\relax
\EndOfBibitem
\bibitem[Rydzewski and Nowak(2017)Rydzewski, and Nowak]{rydzewski2017ligand}
Rydzewski,~J.; Nowak,~W. {Ligand Diffusion in Proteins via Enhanced Sampling in Molecular Dynamics}. \emph{Phys. Life Rev.} \textbf{2017}, \emph{22}, 58--74\relax
\mciteBstWouldAddEndPuncttrue
\mciteSetBstMidEndSepPunct{\mcitedefaultmidpunct}
{\mcitedefaultendpunct}{\mcitedefaultseppunct}\relax
\EndOfBibitem
\bibitem[Valsson \latin{et~al.}(2016)Valsson, Tiwary, and Parrinello]{valsson2016enhancing}
Valsson,~O.; Tiwary,~P.; Parrinello,~M. {Enhancing Important Fluctuations: Rare Events and Metadynamics from a Conceptual Viewpoint}. \emph{Annu. Rev. Phys. Chem.} \textbf{2016}, \emph{67}, 159--184\relax
\mciteBstWouldAddEndPuncttrue
\mciteSetBstMidEndSepPunct{\mcitedefaultmidpunct}
{\mcitedefaultendpunct}{\mcitedefaultseppunct}\relax
\EndOfBibitem
\bibitem[Bussi and Laio(2020)Bussi, and Laio]{bussi2020using}
Bussi,~G.; Laio,~A. {Using Metadynamics to Explore Complex Free-Energy Landscapes}. \emph{Nat. Rev. Phys.} \textbf{2020}, \emph{2}, 200--2012\relax
\mciteBstWouldAddEndPuncttrue
\mciteSetBstMidEndSepPunct{\mcitedefaultmidpunct}
{\mcitedefaultendpunct}{\mcitedefaultseppunct}\relax
\EndOfBibitem
\bibitem[Laio and Parrinello(2002)Laio, and Parrinello]{laio2002escaping}
Laio,~A.; Parrinello,~M. {Escaping Free-Energy Minima}. \emph{Proc. Natl. Acad. Sci. U.S.A.} \textbf{2002}, \emph{99}, 12562--12566\relax
\mciteBstWouldAddEndPuncttrue
\mciteSetBstMidEndSepPunct{\mcitedefaultmidpunct}
{\mcitedefaultendpunct}{\mcitedefaultseppunct}\relax
\EndOfBibitem
\bibitem[Barducci \latin{et~al.}(2008)Barducci, Bussi, and Parrinello]{barducci2008well}
Barducci,~A.; Bussi,~G.; Parrinello,~M. {Well-Tempered Metadynamics: A Smoothly Converging and Tunable Free-Energy Method}. \emph{Phys. Rev. Lett.} \textbf{2008}, \emph{100}, 020603\relax
\mciteBstWouldAddEndPuncttrue
\mciteSetBstMidEndSepPunct{\mcitedefaultmidpunct}
{\mcitedefaultendpunct}{\mcitedefaultseppunct}\relax
\EndOfBibitem
\bibitem[Torrie and Valleau(1977)Torrie, and Valleau]{torrie1977nonphysical}
Torrie,~G.~M.; Valleau,~J.~P. {Nonphysical Sampling Distributions in Monte Carlo Free-Energy Estimation: Umbrella Sampling}. \emph{J. Comp. Phys.} \textbf{1977}, \emph{23}, 187--199\relax
\mciteBstWouldAddEndPuncttrue
\mciteSetBstMidEndSepPunct{\mcitedefaultmidpunct}
{\mcitedefaultendpunct}{\mcitedefaultseppunct}\relax
\EndOfBibitem
\bibitem[K\"{a}stner(2011)]{Kastner2011umbrella}
K\"{a}stner,~J. {Umbrella Sampling}. \emph{Wiley Interdiscip. Rev. Comput. Mol. Sci.} \textbf{2011}, \emph{1}, 932--942\relax
\mciteBstWouldAddEndPuncttrue
\mciteSetBstMidEndSepPunct{\mcitedefaultmidpunct}
{\mcitedefaultendpunct}{\mcitedefaultseppunct}\relax
\EndOfBibitem
\bibitem[Voter(2000)]{voter2000temperature}
Voter,~A.~F. {Temperature-Accelerated Dynamics for Simulation of Infrequent Events}. \emph{J. Chem. Phys.} \textbf{2000}, \emph{112}, 9599--9606\relax
\mciteBstWouldAddEndPuncttrue
\mciteSetBstMidEndSepPunct{\mcitedefaultmidpunct}
{\mcitedefaultendpunct}{\mcitedefaultseppunct}\relax
\EndOfBibitem
\bibitem[Darve \latin{et~al.}(2008)Darve, Rodr{\'\i}guez-G{\'o}mez, and Pohorille]{darve2008adaptive}
Darve,~E.; Rodr{\'\i}guez-G{\'o}mez,~D.; Pohorille,~A. {Adaptive Biasing Force Method for Scalar and Vector Free Energy Calculations}. \emph{J. Chem. Phys.} \textbf{2008}, \emph{128}\relax
\mciteBstWouldAddEndPuncttrue
\mciteSetBstMidEndSepPunct{\mcitedefaultmidpunct}
{\mcitedefaultendpunct}{\mcitedefaultseppunct}\relax
\EndOfBibitem
\bibitem[Fiorin \latin{et~al.}(2013)Fiorin, Klein, and H{\'e}nin]{fiorin2013using}
Fiorin,~G.; Klein,~M.~L.; H{\'e}nin,~J. {Using Collective Variables to Drive Molecular Dynamics Simulations}. \emph{Mol. Phys.} \textbf{2013}, \emph{111}, 3345--3362\relax
\mciteBstWouldAddEndPuncttrue
\mciteSetBstMidEndSepPunct{\mcitedefaultmidpunct}
{\mcitedefaultendpunct}{\mcitedefaultseppunct}\relax
\EndOfBibitem
\bibitem[Noé and Clementi(2017)Noé, and Clementi]{noe2017collective}
Noé,~F.; Clementi,~C. {Collective Variables for the Study of Long-Time Kinetics from Molecular Trajectories: Theory and Methods}. \emph{Curr. Opin. Struct. Biol.} \textbf{2017}, \emph{43}, 141--147\relax
\mciteBstWouldAddEndPuncttrue
\mciteSetBstMidEndSepPunct{\mcitedefaultmidpunct}
{\mcitedefaultendpunct}{\mcitedefaultseppunct}\relax
\EndOfBibitem
\bibitem[Yu \latin{et~al.}(2014)Yu, Chen, Chen, Samanta, Vanden-Eijnden, and Tuckerman]{yu2014order}
Yu,~T.-Q.; Chen,~P.-Y.; Chen,~M.; Samanta,~A.; Vanden-Eijnden,~E.; Tuckerman,~M. Order-Parameter-Aided Temperature-Accelerated Sampling for the Exploration of Crystal Polymorphism and Solid-Liquid Phase Transitions. \emph{J. Chem. Phys.} \textbf{2014}, \emph{140}\relax
\mciteBstWouldAddEndPuncttrue
\mciteSetBstMidEndSepPunct{\mcitedefaultmidpunct}
{\mcitedefaultendpunct}{\mcitedefaultseppunct}\relax
\EndOfBibitem
\bibitem[Tse \latin{et~al.}(2019)Tse, Comer, Sang~Chu, Wang, and Chipot]{tse2019affordable}
Tse,~C.~H.; Comer,~J.; Sang~Chu,~S.~K.; Wang,~Y.; Chipot,~C. Affordable Membrane Permeability Calculations: Permeation of Short-Chain Alcohols through Pure-Lipid Bilayers and a Mammalian Cell Membrane. \emph{J. Chem. Theory Comput.} \textbf{2019}, \emph{15}, 2913--2924\relax
\mciteBstWouldAddEndPuncttrue
\mciteSetBstMidEndSepPunct{\mcitedefaultmidpunct}
{\mcitedefaultendpunct}{\mcitedefaultseppunct}\relax
\EndOfBibitem
\bibitem[Bidon-Chanal \latin{et~al.}(2013)Bidon-Chanal, Krammer, Blot, Pebay-Peyroula, Chipot, Ravaud, and Dehez]{bidon2013membrane}
Bidon-Chanal,~A.; Krammer,~E.-M.; Blot,~D.; Pebay-Peyroula,~E.; Chipot,~C.; Ravaud,~S.; Dehez,~F. How Do Membrane Transporters Sense pH? The Case of the Mitochondrial ADP--ATP Carrier. \emph{J. Phys. Chem. Lett.} \textbf{2013}, \emph{4}, 3787--3791\relax
\mciteBstWouldAddEndPuncttrue
\mciteSetBstMidEndSepPunct{\mcitedefaultmidpunct}
{\mcitedefaultendpunct}{\mcitedefaultseppunct}\relax
\EndOfBibitem
\bibitem[Chipot and H{\'e}nin(2005)Chipot, and H{\'e}nin]{chipot2005exploring}
Chipot,~C.; H{\'e}nin,~J. Exploring the Free-Energy Landscape of a Short Peptide using an Average Force. \emph{J. Chem. Phys.} \textbf{2005}, \emph{123}\relax
\mciteBstWouldAddEndPuncttrue
\mciteSetBstMidEndSepPunct{\mcitedefaultmidpunct}
{\mcitedefaultendpunct}{\mcitedefaultseppunct}\relax
\EndOfBibitem
\bibitem[Bonhenry \latin{et~al.}(2018)Bonhenry, Dehez, and Tarek]{bonhenry2018effects}
Bonhenry,~D.; Dehez,~F.; Tarek,~M. Effects of Hydration on the Protonation State of a Lysine Analog Crossing a Phospholipid Bilayer--Insights from Molecular Dynamics and Free-Energy Calculations. \emph{Phys. Chem. Chem. Phys.} \textbf{2018}, \emph{20}, 9101--9107\relax
\mciteBstWouldAddEndPuncttrue
\mciteSetBstMidEndSepPunct{\mcitedefaultmidpunct}
{\mcitedefaultendpunct}{\mcitedefaultseppunct}\relax
\EndOfBibitem
\bibitem[Samanta \latin{et~al.}(2014)Samanta, Tuckerman, Yu, and E]{samanta2014microscopic}
Samanta,~A.; Tuckerman,~M.~E.; Yu,~T.-Q.; E,~W. Microscopic Mechanisms of Equilibrium Melting of a Solid. \emph{Science} \textbf{2014}, \emph{346}, 729--732\relax
\mciteBstWouldAddEndPuncttrue
\mciteSetBstMidEndSepPunct{\mcitedefaultmidpunct}
{\mcitedefaultendpunct}{\mcitedefaultseppunct}\relax
\EndOfBibitem
\bibitem[Yu and Tuckerman(2011)Yu, and Tuckerman]{yu2011temperature}
Yu,~T.-Q.; Tuckerman,~M.~E. Temperature-Accelerated Method for Exploring Polymorphism in Molecular Crystals Based on Free Energy. \emph{Phys. Rev. Lett.} \textbf{2011}, \emph{107}, 015701\relax
\mciteBstWouldAddEndPuncttrue
\mciteSetBstMidEndSepPunct{\mcitedefaultmidpunct}
{\mcitedefaultendpunct}{\mcitedefaultseppunct}\relax
\EndOfBibitem
\bibitem[Abrams and Vanden-Eijnden(2010)Abrams, and Vanden-Eijnden]{abrams2010large}
Abrams,~C.~F.; Vanden-Eijnden,~E. Large-Scale Conformational Sampling of Proteins using Temperature-Accelerated Molecular Dynamics. \emph{Biophys. J.} \textbf{2010}, \emph{98}, 26a\relax
\mciteBstWouldAddEndPuncttrue
\mciteSetBstMidEndSepPunct{\mcitedefaultmidpunct}
{\mcitedefaultendpunct}{\mcitedefaultseppunct}\relax
\EndOfBibitem
\bibitem[Barducci \latin{et~al.}(2011)Barducci, Bonomi, and Parrinello]{barducci2011metadynamics}
Barducci,~A.; Bonomi,~M.; Parrinello,~M. Metadynamics. \emph{Wiley Interdiscip. Rev. Comput. Mol. Sci.} \textbf{2011}, \emph{1}, 826--843\relax
\mciteBstWouldAddEndPuncttrue
\mciteSetBstMidEndSepPunct{\mcitedefaultmidpunct}
{\mcitedefaultendpunct}{\mcitedefaultseppunct}\relax
\EndOfBibitem
\bibitem[Laio and Gervasio(2008)Laio, and Gervasio]{laio2008metadynamics}
Laio,~A.; Gervasio,~F.~L. Metadynamics: A Method to Simulate Rare Events and Reconstruct the Free Energy in Biophysics, Chemistry and Material Science. \emph{Rep. Prog. Phys.} \textbf{2008}, \emph{71}, 126601\relax
\mciteBstWouldAddEndPuncttrue
\mciteSetBstMidEndSepPunct{\mcitedefaultmidpunct}
{\mcitedefaultendpunct}{\mcitedefaultseppunct}\relax
\EndOfBibitem
\bibitem[Singh \latin{et~al.}(2012)Singh, Chopra, and de~Pablo]{singh2012density}
Singh,~S.; Chopra,~M.; de~Pablo,~J.~J. Density of States--Based Molecular Simulations. \emph{Annu. Rev. Chem. Biomol. Eng.} \textbf{2012}, \emph{3}, 369--394\relax
\mciteBstWouldAddEndPuncttrue
\mciteSetBstMidEndSepPunct{\mcitedefaultmidpunct}
{\mcitedefaultendpunct}{\mcitedefaultseppunct}\relax
\EndOfBibitem
\bibitem[Pietrucci and Laio(2009)Pietrucci, and Laio]{pietrucci2009collective}
Pietrucci,~F.; Laio,~A. A collective Variable for the Efficient Exploration of Protein Beta-Sheet Structures: Application to SH3 and GB1. \emph{J. Chem. Theory Comput.} \textbf{2009}, \emph{5}, 2197--2201\relax
\mciteBstWouldAddEndPuncttrue
\mciteSetBstMidEndSepPunct{\mcitedefaultmidpunct}
{\mcitedefaultendpunct}{\mcitedefaultseppunct}\relax
\EndOfBibitem
\bibitem[Mendels \latin{et~al.}(2018)Mendels, Piccini, and Parrinello]{mendels2018collective}
Mendels,~D.; Piccini,~G.; Parrinello,~M. {Collective Variables from Local Fluctuations}. \emph{J. Phys. Chem. Lett.} \textbf{2018}, \emph{9}, 2776--2781\relax
\mciteBstWouldAddEndPuncttrue
\mciteSetBstMidEndSepPunct{\mcitedefaultmidpunct}
{\mcitedefaultendpunct}{\mcitedefaultseppunct}\relax
\EndOfBibitem
\bibitem[Rogal(2021)]{rogal2021reaction}
Rogal,~J. {Reaction Coordinates in Complex Systems -- A Perspective}. \emph{Eur. Phys. J. B} \textbf{2021}, \emph{94}, 1--9\relax
\mciteBstWouldAddEndPuncttrue
\mciteSetBstMidEndSepPunct{\mcitedefaultmidpunct}
{\mcitedefaultendpunct}{\mcitedefaultseppunct}\relax
\EndOfBibitem
\bibitem[Bussi and Branduardi(2015)Bussi, and Branduardi]{bussi2015free}
Bussi,~G.; Branduardi,~D. Free-Energy Calculations with Metadynamics: Theory and Practice. \emph{Rev. Comput. Chem. Volume 28} \textbf{2015}, 1--49\relax
\mciteBstWouldAddEndPuncttrue
\mciteSetBstMidEndSepPunct{\mcitedefaultmidpunct}
{\mcitedefaultendpunct}{\mcitedefaultseppunct}\relax
\EndOfBibitem
\bibitem[Zheng \latin{et~al.}(2008)Zheng, Chen, and Yang]{zheng2008random}
Zheng,~L.; Chen,~M.; Yang,~W. {Random Walk in Orthogonal Space to Achieve Efficient Free-Energy Simulation of Complex Systems}. \emph{Proc. Natl. Acad. Sci. U.S.A.} \textbf{2008}, \emph{105}, 20227--20232\relax
\mciteBstWouldAddEndPuncttrue
\mciteSetBstMidEndSepPunct{\mcitedefaultmidpunct}
{\mcitedefaultendpunct}{\mcitedefaultseppunct}\relax
\EndOfBibitem
\bibitem[Zhang and Chen(2018)Zhang, and Chen]{zhang2018unfolding}
Zhang,~J.; Chen,~M. {Unfolding Hidden Barriers by Active Enhanced Sampling}. \emph{Phys. Rev. Lett.} \textbf{2018}, \emph{121}, 010601\relax
\mciteBstWouldAddEndPuncttrue
\mciteSetBstMidEndSepPunct{\mcitedefaultmidpunct}
{\mcitedefaultendpunct}{\mcitedefaultseppunct}\relax
\EndOfBibitem
\bibitem[Bonati \latin{et~al.}(2021)Bonati, Piccini, and Parrinello]{bonati2021deep}
Bonati,~L.; Piccini,~G.; Parrinello,~M. Deep Learning the Slow Modes for Rare Events Sampling. \emph{Proc. Natl. Acad. Sci. U.S.A.} \textbf{2021}, \emph{118}, e2113533118\relax
\mciteBstWouldAddEndPuncttrue
\mciteSetBstMidEndSepPunct{\mcitedefaultmidpunct}
{\mcitedefaultendpunct}{\mcitedefaultseppunct}\relax
\EndOfBibitem
\bibitem[Trizio and Parrinello(2021)Trizio, and Parrinello]{trizio2021enhanced}
Trizio,~E.; Parrinello,~M. From Enhanced Sampling to Reaction Profiles. \emph{J. Phys. Chem. Lett.} \textbf{2021}, \emph{12}, 8621--8626\relax
\mciteBstWouldAddEndPuncttrue
\mciteSetBstMidEndSepPunct{\mcitedefaultmidpunct}
{\mcitedefaultendpunct}{\mcitedefaultseppunct}\relax
\EndOfBibitem
\bibitem[Sidky \latin{et~al.}(2020)Sidky, Chen, and Ferguson]{sidky2020machine}
Sidky,~H.; Chen,~W.; Ferguson,~A.~L. Machine Learning for Collective Variable Discovery and Enhanced Sampling in Biomolecular Simulation. \emph{Mol. Phys.} \textbf{2020}, \emph{118}, e1737742\relax
\mciteBstWouldAddEndPuncttrue
\mciteSetBstMidEndSepPunct{\mcitedefaultmidpunct}
{\mcitedefaultendpunct}{\mcitedefaultseppunct}\relax
\EndOfBibitem
\bibitem[Chen(2021)]{chen2021collective}
Chen,~M. {Collective Variable-Based Enhanced Sampling and Machine Learning}. \emph{Eur. Phys. J. B} \textbf{2021}, \emph{94}, 1--17\relax
\mciteBstWouldAddEndPuncttrue
\mciteSetBstMidEndSepPunct{\mcitedefaultmidpunct}
{\mcitedefaultendpunct}{\mcitedefaultseppunct}\relax
\EndOfBibitem
\bibitem[Bonati \latin{et~al.}(2023)Bonati, Trizio, Rizzi, and Parrinello]{bonati2023unified}
Bonati,~L.; Trizio,~E.; Rizzi,~A.; Parrinello,~M. A Unified Framework for Machine Learning Collective Variables for Enhanced Sampling Simulations: mlcolvar. \emph{J. Chem. Phys.} \textbf{2023}, \emph{159}\relax
\mciteBstWouldAddEndPuncttrue
\mciteSetBstMidEndSepPunct{\mcitedefaultmidpunct}
{\mcitedefaultendpunct}{\mcitedefaultseppunct}\relax
\EndOfBibitem
\bibitem[Bonati \latin{et~al.}(2020)Bonati, Rizzi, and Parrinello]{bonati2020data}
Bonati,~L.; Rizzi,~V.; Parrinello,~M. {Data-Driven Collective Variables for Enhanced Sampling}. \emph{J. Phys. Chem. Lett.} \textbf{2020}, \emph{11}, 2998--3004\relax
\mciteBstWouldAddEndPuncttrue
\mciteSetBstMidEndSepPunct{\mcitedefaultmidpunct}
{\mcitedefaultendpunct}{\mcitedefaultseppunct}\relax
\EndOfBibitem
\bibitem[Sipka \latin{et~al.}(2023)Sipka, Erlebach, and Grajciar]{sipka2023constructing}
Sipka,~M.; Erlebach,~A.; Grajciar,~L. Constructing collective variables using invariant learned representations. \emph{J. Chem. Theory Comput.} \textbf{2023}, \emph{19}, 887--901\relax
\mciteBstWouldAddEndPuncttrue
\mciteSetBstMidEndSepPunct{\mcitedefaultmidpunct}
{\mcitedefaultendpunct}{\mcitedefaultseppunct}\relax
\EndOfBibitem
\bibitem[Rydzewski(2023)]{rydzewski2023spectral}
Rydzewski,~J. {Spectral Map: Embedding Slow Kinetics in Collective Variables}. \emph{J. Phys. Chem. Lett.} \textbf{2023}, \emph{14}, 5216--5220\relax
\mciteBstWouldAddEndPuncttrue
\mciteSetBstMidEndSepPunct{\mcitedefaultmidpunct}
{\mcitedefaultendpunct}{\mcitedefaultseppunct}\relax
\EndOfBibitem
\bibitem[Rydzewski(2024)]{rydzewski2024tse}
Rydzewski,~J. {Spectral Map for Slow Collective Variables, Markovian Dynamics, and Transition State Ensembles}. \emph{J. Chem. Theory Comput.} \textbf{2024}, \emph{20}, 7775--7784\relax
\mciteBstWouldAddEndPuncttrue
\mciteSetBstMidEndSepPunct{\mcitedefaultmidpunct}
{\mcitedefaultendpunct}{\mcitedefaultseppunct}\relax
\EndOfBibitem
\bibitem[Dorfer \latin{et~al.}(2015)Dorfer, Kelz, and Widmer]{dorfer2015deep}
Dorfer,~M.; Kelz,~R.; Widmer,~G. Deep Linear Discriminant Analysis. \emph{arXiv preprint arXiv:1511.04707} \textbf{2015}, \relax
\mciteBstWouldAddEndPunctfalse
\mciteSetBstMidEndSepPunct{\mcitedefaultmidpunct}
{}{\mcitedefaultseppunct}\relax
\EndOfBibitem
\bibitem[Lemke and Peter(2019)Lemke, and Peter]{lemke2019encodermap}
Lemke,~T.; Peter,~C. Encodermap: Dimensionality Reduction and Generation of Molecule Conformations. \emph{J. Chem. Theory Comput.} \textbf{2019}, \emph{15}, 1209--1215\relax
\mciteBstWouldAddEndPuncttrue
\mciteSetBstMidEndSepPunct{\mcitedefaultmidpunct}
{\mcitedefaultendpunct}{\mcitedefaultseppunct}\relax
\EndOfBibitem
\bibitem[Sch{\"o}berl \latin{et~al.}(2019)Sch{\"o}berl, Zabaras, and Koutsourelakis]{schoberl2019predictive}
Sch{\"o}berl,~M.; Zabaras,~N.; Koutsourelakis,~P.-S. Predictive Collective Variable Discovery with Deep Bayesian Models. \emph{J. Chem. Phys.} \textbf{2019}, \emph{150}\relax
\mciteBstWouldAddEndPuncttrue
\mciteSetBstMidEndSepPunct{\mcitedefaultmidpunct}
{\mcitedefaultendpunct}{\mcitedefaultseppunct}\relax
\EndOfBibitem
\bibitem[Hern{\'a}ndez \latin{et~al.}(2018)Hern{\'a}ndez, Wayment-Steele, Sultan, Husic, and Pande]{hernandez2018variational}
Hern{\'a}ndez,~C.~X.; Wayment-Steele,~H.~K.; Sultan,~M.~M.; Husic,~B.~E.; Pande,~V.~S. Variational Encoding of Complex Dynamics. \emph{Phys. Rev. E} \textbf{2018}, \emph{97}, 062412\relax
\mciteBstWouldAddEndPuncttrue
\mciteSetBstMidEndSepPunct{\mcitedefaultmidpunct}
{\mcitedefaultendpunct}{\mcitedefaultseppunct}\relax
\EndOfBibitem
\bibitem[Tajs \latin{et~al.}(2025)Tajs, Skarupski, and Rydzewski]{tajs2025neuraltsne}
Tajs,~P.; Skarupski,~M.; Rydzewski,~J. {NeuralTSNE: A Python Package for the Dimensionality Reduction of Molecular Dynamics Data Using Neural Networks}. \emph{J. Chem. Inf. Model.} \textbf{2025}, \relax
\mciteBstWouldAddEndPunctfalse
\mciteSetBstMidEndSepPunct{\mcitedefaultmidpunct}
{}{\mcitedefaultseppunct}\relax
\EndOfBibitem
\bibitem[Bolhuis \latin{et~al.}(2002)Bolhuis, Chandler, Dellago, and Geissler]{bolhuis2002transition}
Bolhuis,~P.~G.; Chandler,~D.; Dellago,~C.; Geissler,~P.~L. {Transition Path Sampling: Throwing Ropes over Rough Mountain Passes, in the Dark}. \emph{Annu. Rev. Phys. Chem.} \textbf{2002}, \emph{53}, 291--318\relax
\mciteBstWouldAddEndPuncttrue
\mciteSetBstMidEndSepPunct{\mcitedefaultmidpunct}
{\mcitedefaultendpunct}{\mcitedefaultseppunct}\relax
\EndOfBibitem
\bibitem[Chodera and Noé(2014)Chodera, and Noé]{chodera2014markov}
Chodera,~J.~D.; Noé,~F. {Markov State Models of Biomolecular Conformational Dynamics}. \emph{Curr. Opin. Struct. Biol.} \textbf{2014}, \emph{25}, 135--144\relax
\mciteBstWouldAddEndPuncttrue
\mciteSetBstMidEndSepPunct{\mcitedefaultmidpunct}
{\mcitedefaultendpunct}{\mcitedefaultseppunct}\relax
\EndOfBibitem
\bibitem[Elber(2020)]{elber2020milestoning}
Elber,~R. Milestoning: An Efficient Approach for Atomically Detailed Simulations of Kinetics in Biophysics. \emph{Annu. Rev. Biophys.} \textbf{2020}, \emph{49}, 69--85\relax
\mciteBstWouldAddEndPuncttrue
\mciteSetBstMidEndSepPunct{\mcitedefaultmidpunct}
{\mcitedefaultendpunct}{\mcitedefaultseppunct}\relax
\EndOfBibitem
\bibitem[Tiwary and Berne(2016)Tiwary, and Berne]{tiwary2016spectral}
Tiwary,~P.; Berne,~B.~J. {Spectral Gap Optimization of Order Parameters for Sampling Complex Molecular Systems}. \emph{Proc. Natl. Acad. Sci. U.S.A.} \textbf{2016}, \emph{113}, 2839\relax
\mciteBstWouldAddEndPuncttrue
\mciteSetBstMidEndSepPunct{\mcitedefaultmidpunct}
{\mcitedefaultendpunct}{\mcitedefaultseppunct}\relax
\EndOfBibitem
\bibitem[Tsai \latin{et~al.}(2020)Tsai, Kuo, and Tiwary]{tsai2020learning}
Tsai,~S.-T.; Kuo,~E.-J.; Tiwary,~P. Learning Molecular Dynamics with Simple Language Model Built upon Long Short-Term Memory Neural Network. \emph{Nat. Commun.} \textbf{2020}, \emph{11}, 5115\relax
\mciteBstWouldAddEndPuncttrue
\mciteSetBstMidEndSepPunct{\mcitedefaultmidpunct}
{\mcitedefaultendpunct}{\mcitedefaultseppunct}\relax
\EndOfBibitem
\bibitem[Rydzewski and G{\"o}kdemir(2024)Rydzewski, and G{\"o}kdemir]{rydzewski2024learning}
Rydzewski,~J.; G{\"o}kdemir,~T. {Learning Markovian Dynamics with Spectral Maps}. \emph{J. Chem. Phys.} \textbf{2024}, \emph{160}, 091102\relax
\mciteBstWouldAddEndPuncttrue
\mciteSetBstMidEndSepPunct{\mcitedefaultmidpunct}
{\mcitedefaultendpunct}{\mcitedefaultseppunct}\relax
\EndOfBibitem
\bibitem[McCarty and Parrinello(2017)McCarty, and Parrinello]{mccarty2017variational}
McCarty,~J.; Parrinello,~M. {A Variational Conformational Dynamics Approach to the Selection of Collective Variables in Metadynamics}. \emph{J. Chem. Phys.} \textbf{2017}, \emph{147}, 204109\relax
\mciteBstWouldAddEndPuncttrue
\mciteSetBstMidEndSepPunct{\mcitedefaultmidpunct}
{\mcitedefaultendpunct}{\mcitedefaultseppunct}\relax
\EndOfBibitem
\bibitem[Belkacemi \latin{et~al.}(2021)Belkacemi, Gkeka, Leli{\`e}vre, and Stoltz]{belkacemi2021chasing}
Belkacemi,~Z.; Gkeka,~P.; Leli{\`e}vre,~T.; Stoltz,~G. {Chasing Collective Variables using Autoencoders and Biased Trajectories}. \emph{J. Chem. Theory Comput.} \textbf{2021}, \emph{18}, 59--78\relax
\mciteBstWouldAddEndPuncttrue
\mciteSetBstMidEndSepPunct{\mcitedefaultmidpunct}
{\mcitedefaultendpunct}{\mcitedefaultseppunct}\relax
\EndOfBibitem
\bibitem[Rydzewski and Valsson(2021)Rydzewski, and Valsson]{rydzewski2021multiscale}
Rydzewski,~J.; Valsson,~O. {Multiscale Reweighted Stochastic Embedding: Deep Learning of Collective Variables for Enhanced Sampling}. \emph{J. Phys. Chem. A} \textbf{2021}, \emph{125}, 6286--6302\relax
\mciteBstWouldAddEndPuncttrue
\mciteSetBstMidEndSepPunct{\mcitedefaultmidpunct}
{\mcitedefaultendpunct}{\mcitedefaultseppunct}\relax
\EndOfBibitem
\bibitem[Rydzewski \latin{et~al.}(2022)Rydzewski, Chen, Ghosh, and Valsson]{rydzewski2022reweighted}
Rydzewski,~J.; Chen,~M.; Ghosh,~T.~K.; Valsson,~O. {Reweighted Manifold Learning of Collective Variables from Enhanced Sampling Simulations}. \emph{J. Chem. Theory Comput.} \textbf{2022}, \emph{18}, 7179--7192\relax
\mciteBstWouldAddEndPuncttrue
\mciteSetBstMidEndSepPunct{\mcitedefaultmidpunct}
{\mcitedefaultendpunct}{\mcitedefaultseppunct}\relax
\EndOfBibitem
\bibitem[Donati \latin{et~al.}(2017)Donati, Hartmann, and Keller]{donati2017girsanov}
Donati,~L.; Hartmann,~C.; Keller,~B.~G. {Girsanov Reweighting for Path Ensembles and Markov State Models}. \emph{J. Chem. Phys.} \textbf{2017}, \emph{146}, 244112\relax
\mciteBstWouldAddEndPuncttrue
\mciteSetBstMidEndSepPunct{\mcitedefaultmidpunct}
{\mcitedefaultendpunct}{\mcitedefaultseppunct}\relax
\EndOfBibitem
\bibitem[Coifman and Lafon(2006)Coifman, and Lafon]{coifman2006diffusion}
Coifman,~R.~R.; Lafon,~S. {Diffusion Maps}. \emph{Appl. Comput. Harmon. Anal.} \textbf{2006}, \emph{21}, 5--30\relax
\mciteBstWouldAddEndPuncttrue
\mciteSetBstMidEndSepPunct{\mcitedefaultmidpunct}
{\mcitedefaultendpunct}{\mcitedefaultseppunct}\relax
\EndOfBibitem
\bibitem[Coifman \latin{et~al.}(2008)Coifman, Kevrekidis, Lafon, Maggioni, and Nadler]{coifman2008diffusion}
Coifman,~R.~R.; Kevrekidis,~I.~G.; Lafon,~S.; Maggioni,~M.; Nadler,~B. {Diffusion Maps, Reduction Coordinates, and Low Dimensional Representation of Stochastic Systems}. \emph{Multiscale Model. Simul.} \textbf{2008}, \emph{7}, 842--864\relax
\mciteBstWouldAddEndPuncttrue
\mciteSetBstMidEndSepPunct{\mcitedefaultmidpunct}
{\mcitedefaultendpunct}{\mcitedefaultseppunct}\relax
\EndOfBibitem
\bibitem[Rohrdanz \latin{et~al.}(2011)Rohrdanz, Zheng, Maggioni, and Clementi]{rohrdanz2011determination}
Rohrdanz,~M.~A.; Zheng,~W.; Maggioni,~M.; Clementi,~C. {Determination of Reaction Coordinates via Locally Scaled Diffusion Map}. \emph{J. Chem. Phys.} \textbf{2011}, \emph{134}, 03B624\relax
\mciteBstWouldAddEndPuncttrue
\mciteSetBstMidEndSepPunct{\mcitedefaultmidpunct}
{\mcitedefaultendpunct}{\mcitedefaultseppunct}\relax
\EndOfBibitem
\bibitem[Rydzewski(2023)]{rydzewski2023selecting}
Rydzewski,~J. {Selecting High-Dimensional Representations of Physical Systems by Reweighted Diffusion Maps}. \emph{J. Phys. Chem. Lett.} \textbf{2023}, \emph{14}, 2778--2783\relax
\mciteBstWouldAddEndPuncttrue
\mciteSetBstMidEndSepPunct{\mcitedefaultmidpunct}
{\mcitedefaultendpunct}{\mcitedefaultseppunct}\relax
\EndOfBibitem
\bibitem[Singer and Coifman(2008)Singer, and Coifman]{singer2008non}
Singer,~A.; Coifman,~R.~R. {Non-Linear Independent Component Analysis with Diffusion Maps}. \emph{Appl. Comput. Harmon. Anal.} \textbf{2008}, \emph{25}, 226--239\relax
\mciteBstWouldAddEndPuncttrue
\mciteSetBstMidEndSepPunct{\mcitedefaultmidpunct}
{\mcitedefaultendpunct}{\mcitedefaultseppunct}\relax
\EndOfBibitem
\bibitem[Singer \latin{et~al.}(2009)Singer, Erban, Kevrekidis, and Coifman]{singer2009detecting}
Singer,~A.; Erban,~R.; Kevrekidis,~I.~G.; Coifman,~R.~R. {Detecting Intrinsic Slow Variables in Stochastic Dynamical Systems by Anisotropic Diffusion Maps}. \emph{Proc. Natl. Acad. Sci. U.S.A.} \textbf{2009}, \emph{106}, 16090--16095\relax
\mciteBstWouldAddEndPuncttrue
\mciteSetBstMidEndSepPunct{\mcitedefaultmidpunct}
{\mcitedefaultendpunct}{\mcitedefaultseppunct}\relax
\EndOfBibitem
\bibitem[Evans \latin{et~al.}(2022)Evans, Cameron, and Tiwary]{evans2022computing}
Evans,~L.; Cameron,~M.~K.; Tiwary,~P. {Computing Committors via Mahalanobis Diffusion Maps with Enhanced Sampling Data}. \emph{J. Chem. Phys.} \textbf{2022}, \emph{157}, 214107\relax
\mciteBstWouldAddEndPuncttrue
\mciteSetBstMidEndSepPunct{\mcitedefaultmidpunct}
{\mcitedefaultendpunct}{\mcitedefaultseppunct}\relax
\EndOfBibitem
\bibitem[Evans \latin{et~al.}(2023)Evans, Cameron, and Tiwary]{evans2023computing}
Evans,~L.; Cameron,~M.~K.; Tiwary,~P. {Computing Committors in Collective Variables via Mahalanobis Diffusion Maps}. \emph{Appl. Comput. Harmon. Anal.} \textbf{2023}, \emph{64}, 62--101\relax
\mciteBstWouldAddEndPuncttrue
\mciteSetBstMidEndSepPunct{\mcitedefaultmidpunct}
{\mcitedefaultendpunct}{\mcitedefaultseppunct}\relax
\EndOfBibitem
\bibitem[Banisch \latin{et~al.}(2020)Banisch, Trstanova, Bittracher, Klus, and Koltai]{banisch2020diffusion}
Banisch,~R.; Trstanova,~Z.; Bittracher,~A.; Klus,~S.; Koltai,~P. {Diffusion Maps Tailored to Arbitrary Non-Degenerate Itô Processes}. \emph{Appl. Comput. Harmon. Anal.} \textbf{2020}, \emph{48}, 242--265\relax
\mciteBstWouldAddEndPuncttrue
\mciteSetBstMidEndSepPunct{\mcitedefaultmidpunct}
{\mcitedefaultendpunct}{\mcitedefaultseppunct}\relax
\EndOfBibitem
\bibitem[Trstanova \latin{et~al.}(2020)Trstanova, Leimkuhler, and Leli{\`e}vre]{trstanova2020local}
Trstanova,~Z.; Leimkuhler,~B.; Leli{\`e}vre,~T. {Local and Global Perspectives on Diffusion Maps in the Analysis of Molecular Systems}. \emph{Proc. Royal Soc. A} \textbf{2020}, \emph{476}, 20190036\relax
\mciteBstWouldAddEndPuncttrue
\mciteSetBstMidEndSepPunct{\mcitedefaultmidpunct}
{\mcitedefaultendpunct}{\mcitedefaultseppunct}\relax
\EndOfBibitem
\bibitem[Maragliano \latin{et~al.}(2006)Maragliano, Fischer, Vanden-Eijnden, and Ciccotti]{maragliano2006string}
Maragliano,~L.; Fischer,~A.; Vanden-Eijnden,~E.; Ciccotti,~G. {String Method in Collective Variables: Minimum Free Energy Paths and Isocommittor Surfaces}. \emph{J. Chem. Phys.} \textbf{2006}, \emph{125}, 024106\relax
\mciteBstWouldAddEndPuncttrue
\mciteSetBstMidEndSepPunct{\mcitedefaultmidpunct}
{\mcitedefaultendpunct}{\mcitedefaultseppunct}\relax
\EndOfBibitem
\bibitem[Keller and Bolhuis(2024)Keller, and Bolhuis]{keller2024dynamical}
Keller,~B.~G.; Bolhuis,~P.~G. Dynamical Reweighting for Biased Rare Event Simulations. \emph{Annual Review of Physical Chemistry} \textbf{2024}, \emph{75}, 137--162\relax
\mciteBstWouldAddEndPuncttrue
\mciteSetBstMidEndSepPunct{\mcitedefaultmidpunct}
{\mcitedefaultendpunct}{\mcitedefaultseppunct}\relax
\EndOfBibitem
\bibitem[Tiwary and Parrinello(2015)Tiwary, and Parrinello]{tiwary2015time}
Tiwary,~P.; Parrinello,~M. {A Time-Independent Free Energy Estimator for Metadynamics}. \emph{J. Phys. Chem. B} \textbf{2015}, \emph{119}, 736--742\relax
\mciteBstWouldAddEndPuncttrue
\mciteSetBstMidEndSepPunct{\mcitedefaultmidpunct}
{\mcitedefaultendpunct}{\mcitedefaultseppunct}\relax
\EndOfBibitem
\bibitem[Liu \latin{et~al.}(2024)Liu, Ghosh, Lin, and Chen]{liu2024unbiasing}
Liu,~Y.; Ghosh,~T.~K.; Lin,~G.; Chen,~M. {Unbiasing Enhanced Sampling on a High-Dimensional Free Energy Surface with a Deep Generative Model}. \emph{J. Phys. Chem. Lett.} \textbf{2024}, \emph{15}, 3938--3945\relax
\mciteBstWouldAddEndPuncttrue
\mciteSetBstMidEndSepPunct{\mcitedefaultmidpunct}
{\mcitedefaultendpunct}{\mcitedefaultseppunct}\relax
\EndOfBibitem
\bibitem[Rydzewski \latin{et~al.}(2023)Rydzewski, Chen, and Valsson]{rydzewski2023manifold}
Rydzewski,~J.; Chen,~M.; Valsson,~O. {Manifold Learning in Atomistic Simulations: A Conceptual Review}. \emph{Mach. Learn.: Sci. Technol.} \textbf{2023}, \emph{4}, 031001\relax
\mciteBstWouldAddEndPuncttrue
\mciteSetBstMidEndSepPunct{\mcitedefaultmidpunct}
{\mcitedefaultendpunct}{\mcitedefaultseppunct}\relax
\EndOfBibitem
\bibitem[van~der Maaten and Hinton(2008)van~der Maaten, and Hinton]{maaten2008visualizing}
van~der Maaten,~L.; Hinton,~G. {Visualizing Data using $t$-SNE}. \emph{J. Mach. Learn. Res.} \textbf{2008}, \emph{9}, 2579--2605\relax
\mciteBstWouldAddEndPuncttrue
\mciteSetBstMidEndSepPunct{\mcitedefaultmidpunct}
{\mcitedefaultendpunct}{\mcitedefaultseppunct}\relax
\EndOfBibitem
\bibitem[van~der Maaten(2009)]{maaten2009learning}
van~der Maaten,~L. {Learning a Parametric Embedding by Preserving Local Structure}. \emph{J. Mach. Learn. Res.} \textbf{2009}, \emph{5}, 384--391\relax
\mciteBstWouldAddEndPuncttrue
\mciteSetBstMidEndSepPunct{\mcitedefaultmidpunct}
{\mcitedefaultendpunct}{\mcitedefaultseppunct}\relax
\EndOfBibitem
\bibitem[Giberti \latin{et~al.}(2020)Giberti, Cheng, Tribello, and Ceriotti]{giberti2020iterative}
Giberti,~F.; Cheng,~B.; Tribello,~G.~A.; Ceriotti,~M. {Iterative Unbiasing of Quasi-Equilibrium Sampling}. \emph{J. Chem. Theory Comput.} \textbf{2020}, \emph{16}, 100--107\relax
\mciteBstWouldAddEndPuncttrue
\mciteSetBstMidEndSepPunct{\mcitedefaultmidpunct}
{\mcitedefaultendpunct}{\mcitedefaultseppunct}\relax
\EndOfBibitem
\bibitem[Sanbonmatsu and Garc{\'\i}a(2002)Sanbonmatsu, and Garc{\'\i}a]{sanbonmatsu2002structure}
Sanbonmatsu,~K.~Y.; Garc{\'\i}a,~A.~E. Structure of Met-enkephalin in Explicit Aqueous Solution using Replica Exchange Molecular Dynamics. \emph{Proteins} \textbf{2002}, \emph{46}, 225--234\relax
\mciteBstWouldAddEndPuncttrue
\mciteSetBstMidEndSepPunct{\mcitedefaultmidpunct}
{\mcitedefaultendpunct}{\mcitedefaultseppunct}\relax
\EndOfBibitem
\bibitem[Evans and Wales(2003)Evans, and Wales]{evans2003free}
Evans,~D.~A.; Wales,~D.~J. The Free Energy Landscape and Dynamics of Met-enkephalin. \emph{J. Chem. Phys.} \textbf{2003}, \emph{119}, 9947--9955\relax
\mciteBstWouldAddEndPuncttrue
\mciteSetBstMidEndSepPunct{\mcitedefaultmidpunct}
{\mcitedefaultendpunct}{\mcitedefaultseppunct}\relax
\EndOfBibitem
\bibitem[Abrams and Tuckerman(2008)Abrams, and Tuckerman]{abrams2008efficient}
Abrams,~J.~B.; Tuckerman,~M.~E. Efficient and Direct Generation of Multidimensional Free Energy Surfaces via Adiabatic Dynamics without Coordinate Transformations. \emph{J. Phys. Chem. B} \textbf{2008}, \emph{112}, 15742--15757\relax
\mciteBstWouldAddEndPuncttrue
\mciteSetBstMidEndSepPunct{\mcitedefaultmidpunct}
{\mcitedefaultendpunct}{\mcitedefaultseppunct}\relax
\EndOfBibitem
\bibitem[Rosso \latin{et~al.}(2002)Rosso, Min{\'a}ry, Zhu, and Tuckerman]{rosso2002use}
Rosso,~L.; Min{\'a}ry,~P.; Zhu,~Z.; Tuckerman,~M.~E. On the Use of the Adiabatic Molecular Dynamics Technique in the Calculation of Free Energy Profiles. \emph{J. Chem. Phys.} \textbf{2002}, \emph{116}, 4389--4402\relax
\mciteBstWouldAddEndPuncttrue
\mciteSetBstMidEndSepPunct{\mcitedefaultmidpunct}
{\mcitedefaultendpunct}{\mcitedefaultseppunct}\relax
\EndOfBibitem
\bibitem[Maragliano and Vanden-Eijnden(2006)Maragliano, and Vanden-Eijnden]{maragliano2006temperature}
Maragliano,~L.; Vanden-Eijnden,~E. A Temperature Accelerated Method for Sampling Free Energy and Determining Reaction Pathways in Rare Events Simulations. \emph{Chem. Phys. Lett.} \textbf{2006}, \emph{426}, 168--175\relax
\mciteBstWouldAddEndPuncttrue
\mciteSetBstMidEndSepPunct{\mcitedefaultmidpunct}
{\mcitedefaultendpunct}{\mcitedefaultseppunct}\relax
\EndOfBibitem
\end{mcitethebibliography}

\end{document}